\documentclass[aps,
							prc,
							preprint,
							superscriptaddress,
							nofootinbib
							]{revtex4-1}

\usepackage[english]{babel}
\usepackage[ansinew]{inputenc}
\usepackage{microtype}
\usepackage{color,
						bbm,
						slashed,
						SIunits,
						graphicx
						}
\usepackage[fleqn]{amsmath}

\begin{document}

\title{Wilson coefficients and four-quark condensates in QCD sum rules for medium modifications of \texorpdfstring{$\boldsymbol D$}{D} mesons}

\date{\today}

\author{T.\ Buchheim}
 \email{t.buchheim@hzdr.de}
\affiliation{Helmholtz-Zentrum Dresden-Rossendorf, PF 510119, D-01314 Dresden, Germany}
\affiliation{Technische Universit{\"a}t Dresden, Institut f\"{u}r Theoretische Physik, D-01062 Dresden, Germany}

\author{T.\ Hilger}
\affiliation{University of Graz, Institute of Physics, NAWI Graz, A-8010 Graz, Austria}

\author{B.\ K{\"a}mpfer}
\affiliation{Helmholtz-Zentrum Dresden-Rossendorf, PF 510119, D-01314 Dresden, Germany}
\affiliation{Technische Universit{\"a}t Dresden, Institut f\"{u}r Theoretische Physik, D-01062 Dresden, Germany}

\begin{abstract}
Wilson coefficients of light four-quark condensates in QCD sum rules are evaluated for pseudo-scalar $D$ mesons, thus, pushing the sum rules toward mass dimension six. Contrary to the situation for $\bar{q}q$ mesons the impact of the four-quark condensates for vacuum as well as in-medium situations is found to be rather small within the Borel window used in previous analyses. The complete four-quark condensate contributions enable to identify candidates for an order parameter of spontaneous chiral symmetry breaking/restoration as well as to evaluate stability criteria of operator product expansions.
\end{abstract}



\maketitle

\section{Introduction}
\label{sec:intro}

QCD sum rules \cite{svz79} represent a valuable tool for establishing a link of Quantum ChromoDynamics (QCD), formulated in quark and gluon degrees of freedom, and hadron physics. By separating the soft, long-range phenomena and the hard, perturbatively calculable effects many hadronic properties become accessible. Hereby, the condensates, i.\,e.\ expectation values of QCD operators absorbing the long-range effects, serve as input parameters which can be adjusted at selected observables. The separation of scales seems unproblematic in the light-quark sector. However, when including heavy quarks, their masses enter the scheme as additional scales requiring extra effort. Nevertheless, since other methods (such as lattice QCD evaluations, Schr{\"o}dinger equation approaches with potentials, Dyson-Schwinger and Bethe-Salpeter equations, etc.) are at our disposal, a mutual judging is of interest.

Among the central issues of hadron physics in the light-quark sector is chiral symmetry and its breaking pattern. If one relates spontaneous chiral symmetry breaking with the non-zero value of the chiral condensate in vacuum, $\langle\bar{q}q\rangle_{0} \approx (\unit{-245}{MeV})^{3}$, one is tempted to ask for observable consequences of chiral restoration, i.\,e.\ to which extent do hadron observables change under a change of the chiral condensate \cite{weise94}. In leading order, at non-zero temperature $T$ and/or density $n$, the chiral condensate is modified according to $ \langle \bar{q}q \rangle_{T,n} \approx \langle \bar{q}q \rangle_{0} \left( 1 - \frac{T^{2}}{8f^{2}_{\pi}} - \frac{\sigma_\text{N}\, n}{m_{\pi}^{2} f_{\pi}^{2}} \right) $ (cf.~\cite{hatsukoike,cohen95}), 
where non-zero temperatures are modeled by a pion gas and non-zero densities by ambient nucleons. The symbol $f_{\pi}$ denotes the pion decay constant, $m_{\pi}$ the pion mass and $\sigma_\text{N}$ is the nucleon sigma term \cite{cohen95}. That means, at non-zero temperature and/or density the chiral condensate is diminished relative to its vacuum value. Chiral restoration may be understood accordingly as being necessarily accompanied by $\langle \bar{q}q \rangle_{T,n} \rightarrow 0$. Also further condensates, especially four-quark condensates, exhibit non-invariant behavior under chiral transformations \cite{leupold06,thomas07,hilger11}. Such condensates are candidates for order parameters of spontaneous chiral symmetry breaking and restoration similarly to the chiral condensate.

While deconfinement accompanied by dissolving hadron states is the obviously strongest medium modification of hadrons, more modest modifications are envisaged during the last two decennia (cf.\ \cite{hatsukoike,cohen95,jin} for surveys on medium modifications of hadrons). The seminal paper by Hatsuda and Lee \cite{hatsulee} devices a scenario where spectral properties of mesons (most notably condensed into moments characterizing masses and widths) do change in a strongly interacting environment. Clearly, there are further condensates which change at non-zero temperature and density \cite{hatsukoike,jin}. Most prominently, the gluon condensate related to scale invariance breaking exhibits such behavior.

Evaluation of QCD sum rules means the following, within the present context. The hadronic spectral density $\rho(\omega^{2}) = \text{Im}\Pi(\omega^{2})/\pi$ is related to the current-current correlator $\Pi(p)$ by a vacuum dispersion relation $\frac{1}{\pi}\int_{0}^{\infty} d\omega^{2} \, \text{Im}\Pi(\omega^{2}) / (\omega^{2} - p^{2}) = \Pi(p^{2})$ with the r.\,h.\,s.\ being accessible perturbatively at small distances. As a first step, the operator product expansion (OPE), after a Borel transformation $\Pi(p^{2}) \longrightarrow \widehat{\Pi}(M^{2})$ \cite{svz79,rry}, leads to a representation $\widehat{\Pi}(M^{2}) = \sum\limits_{k} C_{k} (M^{2})\, \langle \mathcal{O}_{k} \rangle$, where $\langle \mathcal{O}_{k} \rangle = \left\{ \mathbbm{1}, \langle\bar{q}q\rangle, \langle (\alpha_\text{s}/\pi) G^{2} \rangle, \ldots \right\}$ contains the unit element $\mathbbm{1}$, associated to the perturbative term, and condensates, e.\,g.\ $\langle \mathcal{O}_{1} \rangle = \langle\bar{q}q\rangle$, $\langle \mathcal{O}_{2} \rangle = \langle (\alpha_\text{s}/\pi) G^{2} \rangle$ etc., associated to the non-perturbative terms, dubbed power corrections; $C_{k} (M^{2})$ are the corresponding Wilson coefficients, which depend on the Borel mass $M$ as remainder of the momentum ($p$) dependence of the correlator. In such a way the long-range phenomena are separated from short-range phenomena. For the $\rho$ meson at rest the series reads\footnote{The OPE leads to an asymptotic series, where dots in the displayed series denote higher power corrections which may brake down the expansion, thus, requiring a careful evaluation of the OPE's convergence behavior.}
\begin{align}
	\widehat{\Pi}^{(\rho)}(M^{2}) = C^{(\rho)}_{0} \, M^{2} + \frac{C^{(\rho)}_{1}}{M^{2}} \langle\bar{q}q\rangle + \frac{C^{(\rho)}_{2}}{M^{2}} \langle \frac{\alpha_\text{s}}{\pi} G^{2} \rangle + \frac{C^{(\rho)}_{3}}{M^{4}} \langle \mathcal{O}_{3} \rangle + \ldots \; ,
	\label{eq:OPErho}
\end{align}
where the superscript '$(\rho)$' is a reminder that, for the moment, we are talking about the $\rho$ meson which has been analyzed extensively \cite{CBMBook}.

The second step in the sum rule evaluation consists in deducing properties of $\rho(\omega^{2})$ once $\widehat{\Pi}(M^{2})$ is given. We focus on step one, i.\,e.\ the calculation of the so-called OPE side, e.\,g.\ in the form of the series expansion \eqref{eq:OPErho}. Writing schematically $\langle \mathcal{O}_{3} \rangle = \kappa \langle\bar{q}q\rangle^{2}$ with a fudge factor $\kappa$ one observes in fact that, for Borel masses $M \sim \unit{1}{GeV}$, the chiral condensate term $\propto C^{(\rho)}_{1} / M^{2}$ is numerically suppressed, and the gluon condensate term $\propto C^{(\rho)}_{2} / M^{2}$ as well as the four-quark condensate combinations $\propto C^{(\rho)}_{3} / M^{4}$ are of the same order of magnitude for a typical choice $\kappa \sim 2$  \cite{rappwambachvanhees}:
\begin{align}
	& \widehat{\Pi}^{(\rho)}(M^2) \nonumber\\
	& \quad = \frac{1+\frac{\alpha_\text{s}}{\pi}}{8\pi^2} M^2 \log\frac{\mu^2}{M^2} + \frac{m_q}{M^2} \langle\bar{q}q\rangle \qquad\; + \frac{1}{24M^2}\langle\frac{\alpha_\text{s}}{\pi}G^2\rangle  - \frac{112 \pi\alpha_\text{s}}{81 M^4} \kappa \langle\bar{q}q\rangle^2 + \ldots \nonumber\\
	& \quad = \frac{M^2}{8\pi^2} \left( 1.11 \log\frac{\mu^2}{M^2} - 0.0058\, \frac{\text{GeV}^{\,4}}{M^4} + 0.039\, \frac{\text{GeV}^{\,4}}{M^4} \;\;\, - 0.026\,\kappa\, \frac{\text{GeV}^{\,6}}{M^6} \,\! + \ldots \right) \, ,
	\label{eq:OPErhoNumber}
\end{align}
where $\alpha_\text{s} = g^2/4\pi = 0.35$, $m_q=\unit{0.005}{GeV}$, $\langle\bar{q}q\rangle = (-\unit{0.245}{GeV})^3 $ and $\langle (\alpha_\text{s}/\pi) G^2 \rangle = \unit{0.012}{GeV^4}$ have been used.

In the $qQ$ sector\footnote{We use henceforth the shorthand notation $qQ$ for $\bar{q}Q$ and $\bar{Q}q$ mesons. The correlators of mesons $\bar{q}Q$ and anti-mesons $\bar{Q}q$ satisfy the relation $\Pi_{\bar{q}Q}(p) = \Pi_{\bar{Q}q}(-p)$ \cite{zsch11}.}, the situation is different: The chiral condensate appears in the scale-dependent combination $m_Q\langle\bar{q}q\rangle$, i.\,e.\ the heavy quark mass $m_Q$ acts as an amplification factor of the chiral condensate term with sizable impact on spectral properties of $qQ$ mesons~\cite{hilger11}. Furthermore, the in-medium sum rule has an even and an odd part w.\,r.\,t.\ the meson energy $p_0$, satisfying $\Pi(p) =  \Pi^\text{even} (p_0,\vec{p}) + p_0\, \Pi^\text{odd} (p_0,\vec{p})$, since particles and anti-particles are to be distinguished, i.\,e.\ the above dispersion integral runs now over positive and negative frequencies.
In the light chiral limit, $m_q \rightarrow 0$, the first known terms have the structure \cite{hilger09, zsch11}
\begin{subequations}
\label{eq:Dto5}
\begin{align}
	\widehat{\Pi}^\text{even}(M^2) 	&	= C_0 + e^{-m_Q^2/M^2} \sum_{k=1}^6 c^\text{even}_k(M^2) \, \langle \mathcal{O}_k \rangle^\text{even} \, , 
	\label{eq:Deven}\\
	\widehat{\Pi}^\text{odd}(M^2) 	&	= e^{-m_Q^2/M^2} \sum_{k=1}^3 c^\text{odd}_k(M^2) \, \langle \mathcal{O}_k \rangle^\text{odd}
	\label{eq:Dodd}
\end{align}
\end{subequations}
with the perturbative term $C_0$ and condensates $\langle \mathcal{O}_1 \rangle^\text{even} = \langle \mathcal{O}_1 \rangle$, $\langle \mathcal{O}_2 \rangle^\text{even} = \langle\bar{q}g\sigma G q \rangle$, $\langle \mathcal{O}_3 \rangle^\text{even} = \langle \mathcal{O}_2 \rangle $, $\langle \mathcal{O}_4 \rangle^\text{even} = \langle (\alpha_\text{s}/\pi) \left[ (vG)^2/v^2 - G^2/4 \right] \rangle$, $\langle \mathcal{O}_5 \rangle^\text{even} = \langle q^\dagger i D_0 q\rangle$, $\langle \mathcal{O}_6 \rangle^\text{even} = \langle \bar{q} \left[ D_0^2 - g \sigma G / 8 \right] q\rangle$ and $\langle \mathcal{O}_1 \rangle^\text{odd} = \langle q^\dagger q \rangle$, $\langle \mathcal{O}_2 \rangle^\text{odd} = \langle q^\dagger D_0^2 q \rangle$, $\langle \mathcal{O}_3 \rangle^\text{odd} = \langle q^\dagger g \sigma G q\rangle$. The coefficients $c^\text{even,odd}_k(M^2)$ are the Wilson coefficients modulo a common factor $e^{-m_Q^2/M^2}$.
Without knowledge of the Wilson coefficients of the four-quark condensates it is hardly possible to estimate their impact on $\widehat{\Pi}$ in the sum rule and a simple order-of-magnitude comparison can be misleading. (For example, in the above $\rho$ meson sum rule \eqref{eq:OPErhoNumber}, $\langle (\alpha_\text{s}/\pi) G^2 \rangle = \unit{0.012}{GeV^4}$ and $\langle\bar{q}q\rangle^2 = \unit{0.00022}{GeV^6}$ would one lead to guess that the latter condensate contribution is negligible at $M \sim \unit{1}{GeV}$. However, it is the Wilson coefficient $C^{(\rho)}_2 = 1/24$ which makes the gluon contribution comparable to the four-quark condensate term with $C^{(\rho)}_3 = (112/81) \pi\alpha_\text{s}$.)
Therefore, a calculation of Wilson coefficients of the in-medium four-quark condensates entering QCD sum rules for $qQ$ mesons is mandatory. This is the goal of the present paper. 
Equipped with these four-quark condensate contributions one can extend previous OPE/QCD sum rule studies of spectral properties of pseudo-scalar $qQ$ mesons.
Furthermore, and more importantly, one is able to identify four-quark condensate contributions which are not invariant under chiral transformations and, thus, may serve as order parameters of spontaneous chiral symmetry breaking. As pointed out in Ref.~\cite{hilger11}, also in $qQ$ meson systems, the splitting of the spectral densities between parity partners is driven by such order parameters.

Medium modifications of $D$ mesons have become an interesting topic in recent years, since open charm mesons and charmonium serve as probes of hot nuclear matter and deconfinement effects \cite{CBMBook}. Mesons with charm (or bottom) can serve equally well as probes of dense or even saturated nuclear matter (cf.\ \cite{yasuisudoh14,yasuisudoh13b,yasuisudoh13a,blaschke,he,tolos13a,tolos09,kumarmishra11,kumarmishra10} for recent works and further references). For such theoretical investigations the finite-density QCD sum rules look promising \cite{hay00,hilger09,zsch11,wang11,wang13public}.
Although four-quark condensates can influence the in-medium properties significantly, as recalled above for the $\rho$ meson, no light four-quark condensate contributions of the OPE have been used so far to improve the evaluation of in-medium modifications of $qQ$ mesons. Accordingly, we are going to include here light four-quark condensate contributions, thus, extending the previous studies.

The common problem of presently poorly known numerical values of four-quark condensates requires a further treatment of these contributions.
Thus, we resort here to the argument that heavy quarks are static and do not condense, i.\,e.\ one neglects all contributions of condensates containing heavy-quark operators \cite{rry}. Light four-quark condensates are eventually factorized based on the ground state saturation hypothesis to arrive at an order of magnitude estimate of their impact.

Including four-quark condensates is a difficile task and has only been done in vacuum $qQ$ systems employing factorization up to now. In vacuum, the number of terms involves only a small amount of the possible in-medium contributions. Furthermore, assuming vacuum saturation from the very beginning in order to factorize four-quark condensates simplifies evaluations drastically. This is the reason why even in vacuum only few works deal with non-factorized four-quark condensates \cite{hilger12, hohlerrapp12}, whereby even factorized four-quark condensates have never been considered in $qQ$ systems in the medium.

In this work, the non-factorized light four-quark condensate contributions in the medium are presented, and their numerical impact to the OPE is estimated using the factorization hypothesis.
The presented numerical examples focus on open charm mesons in nuclear matter to be studied at FAIR by the CBM \cite{cbm} and PANDA \cite{panda} collaborations, since such an investigation is mandatory to provide an improved theoretical basis for these future large-scale experiments.

Our paper is organized as follows. In section \ref{sec:QSRforqQ}, we provide the evaluation of the OPE leading to the power corrections of the perturbative terms of the current-current correlator. In doing so, we apply a certain truncation scheme which includes systematically light four-quark condensates in leading-order of the strong coupling. We list the four-quark condensates and their Wilson coefficients for pseudo-scalar $qQ$ mesons in section \ref{sec:wilcoeffncond}. Section \ref{sec:HQEandFac} employs factorization of four-quark condensates to arrive at a sum rule where only known condensates enter. Numerical estimates of four-quark condensate contributions to the in-medium OPE of $qQ$ mesons are presented in section \ref{sec:numeval}. Section \ref{sec:CompRhoD} compares four-quark condensate contributions of $\rho$ and $D$ mesons. The summary can be found in section \ref{sec:disnconcl}. The Appendix \ref{sec:calc4Q} details the herein employed OPE technique, such that all results can be confirmed by the interested reader.

\section{QCD sum rules for \texorpdfstring{$\boldsymbol{qQ}$}{qQ} mesons}
\label{sec:QSRforqQ}

We consider the causal current-current correlator in leading-order perturbation theory $\alpha_s^0$
\begin{align}
	\Pi (p) & = i \int d^4x \; e^{ipx} \!\langle \text{T}\left[ j(x) j^\dagger(0) \right] \rangle
	\label{eq:SPcccorr}
\end{align}
with the interpolating pseudo-scalar current 
\begin{align}
	j(x) = \bar{q}(x) i\gamma_5 Q(x) \, ,
\end{align}
where $\text{T}[\ldots]$ means time ordering and $\langle\ldots\rangle$ denotes the Gibbs average. We compute Wilson coefficients using the background field method in Fock-Schwinger gauge $x^\mu A_\mu(x) = 0$ \cite{nov84}. A compact description of the calculus in leading-order perturbation theory is provided in \cite{zsch11,hilger11}. Utilizing Wick's theorem the correlator decomposes as
\begin{align}
	\Pi(p) & = \Pi^{(0)} (p) + \Pi^{(2)} (p)
	\label{eq:OPEs} \\
				& = \parbox[c][15mm][c]{20mm}{\mbox{\includegraphics[width=2cm]{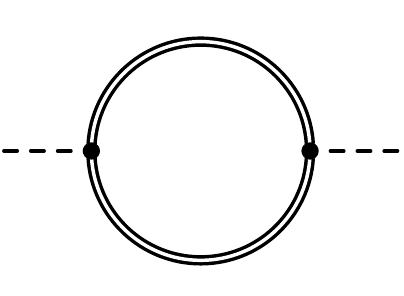}}}
					+ \parbox[c][15mm][c]{20mm}{\mbox{\includegraphics[width=2cm]{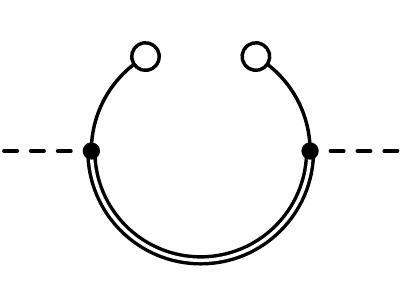}}}	\nonumber
\end{align}
with
\begin{subequations}
\label{eq:PiAlpha0}
\begin{align}
	& \Pi^{(0)} (p) = - i \int d^4x \, e^{ipx} \, \langle : \text{Tr}_\text{c,D}\left[ S_q(0,x) \gamma_5 S_Q(x,0) \gamma_5 \right] : \rangle  \, ,
	\label{eq:fulcontractedterms}	\\
	& \Pi^{(2)} (p) = + \int d^4x \, e^{ipx} \, \langle : \bar{Q}(0) \gamma_5 S_q(0,x) \gamma_5 Q(x) + \bar{q}(x) \gamma_5 S_Q(x,0) \gamma_5 q(0) :\rangle \, ,
	\label{eq:2qterms}
\end{align}
\end{subequations}
where terms associated with disconnected diagrams are omitted. The notation $\text{Tr}_\text{c,D}$ means traces over color and Dirac indices respectively and $: \ldots :$ represents normal ordering w.\,r.\,t.\ the perturbative ground state. $\Pi^{(0)} (p)$ denotes the fully Wick-contracted term and $\Pi^{(2)} (p)$ is the two quark term, i.\,e.\ the superscript number in parentheses refers to the number of Wick-uncontracted quark field operators of the interpolating currents. The full light-quark propagator in the gluonic background field is defined as $S_q(x,y) = -i\langle 0 | \text{T}\left[ q(x) \bar{q}(y) \right] | 0 \rangle$ (and with $q \longrightarrow Q$ for the full heavy-quark propagator, cf.\ the Appendix \ref{subsec:calcsglu} for further details).
\begin{figure}%
\bgroup
\makeatletter\@fleqnfalse\makeatother
\begin{align*}
	\Pi^{(0)}	& = \parbox[c][15mm][c]{20mm}{\mbox{\includegraphics[width=2cm]{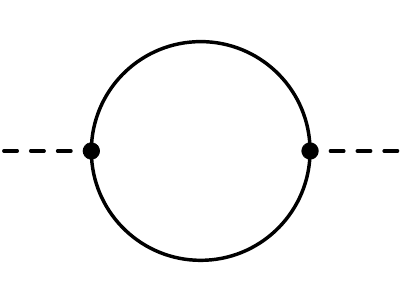}}}^{\!\!\!\!\!\!\text{(a)}\!\!}	
							+ \parbox[c][15mm][c]{20mm}{\mbox{\includegraphics[width=2cm]{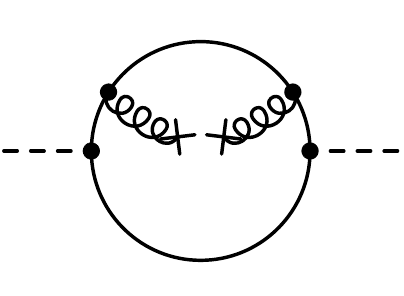}}}^{\!\!\!\!\!\!\text{(b)}\!\!}
							+ \parbox[c][15mm][c]{20mm}{\mbox{\includegraphics[width=2cm]{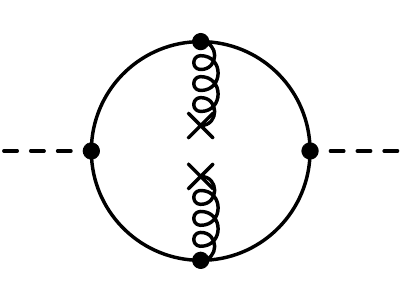}}}^{\!\!\!\!\!\!\text{(c)}\!\!}
							+ \parbox[c][15mm][c]{20mm}{\mbox{\includegraphics[width=2cm]{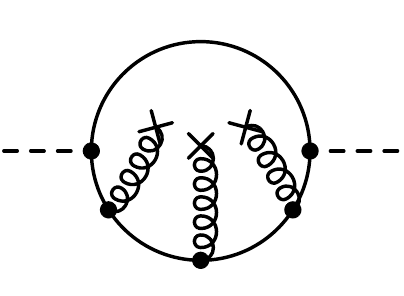}}}^{\!\!\!\!\!\!\text{(d)}\!\!}
							+ \parbox[c][15mm][c]{20mm}{\mbox{\includegraphics[width=2cm]{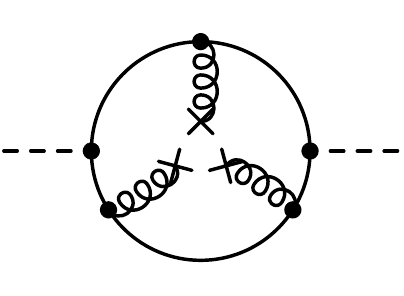}}}^{\!\!\!\!\!\!\text{(e)}\!\!}	\\
						&	+ \parbox[c][15mm][c]{20mm}{\mbox{\includegraphics[width=2cm]{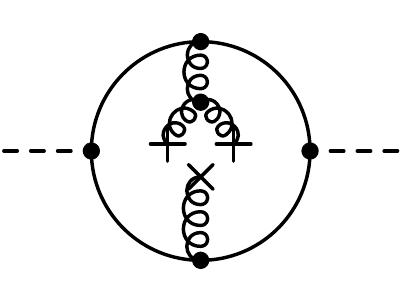}}}^{\!\!\!\!\!\!\text{(f)}\!\!}
							+ \parbox[c][15mm][c]{20mm}{\mbox{\includegraphics[width=2cm]{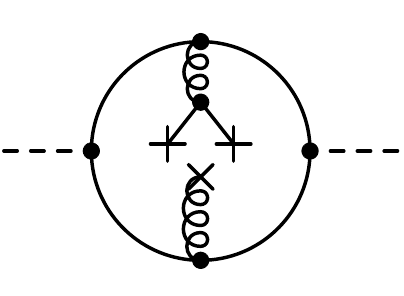}}}^{\!\!\!\!\!\!\text{(g)}\!\!}
							+ \parbox[c][15mm][c]{20mm}{\mbox{\includegraphics[width=2cm]{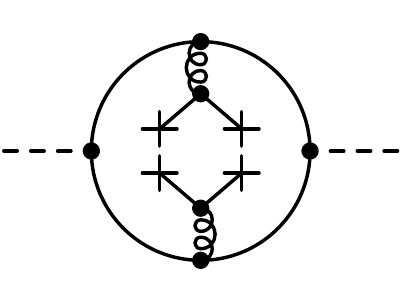}}}^{\!\!\!\!\!\!\text{(h)}\!\!}
							+	\ldots	\nonumber
\end{align*}
\egroup
\caption{Diagrammatic representation of the contribution \eqref{eq:fulcontractedterms} to the current-current correlator, where a selection of topologically relevant diagrams is displayed. Solid lines stand for free quark propagators, wiggly lines are for gluons and crosses symbolize local quark or gluon condensation. In our expansion scheme, we retain only the diagrams (a$-$c) contributing to the perturbative term $C_0$ (a)  and yielding the Wilson coefficients $c^\text{even}_3$ and $c^\text{even}_4$ (b, c) in Eq.~\eqref{eq:Deven} since the other ones (d$-$h) are of higher order in $g$.}%
\label{fig:diagramPiAlpha0(0)}%
\end{figure}
In the diagrammatic representation of the decomposition \eqref{eq:OPEs} dashed lines denote the pseudo-scalar current, double lines symbolize full quark propagators whereas single lines are for free quark propagators, and circles denote non-local quark condensation.

\begin{figure}%
\bgroup
\makeatletter\@fleqnfalse\makeatother
\begin{align*}
	\Pi^{(2)} & = \parbox[c][15mm][c]{20mm}{\mbox{\includegraphics[width=2cm]{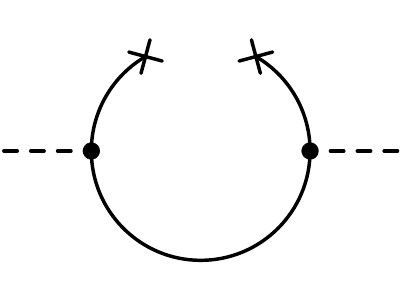}}}^{\!\!\!\!\!\!\text{(a)}\!\!}	
							+ \parbox[c][15mm][c]{20mm}{\mbox{\includegraphics[width=2cm]{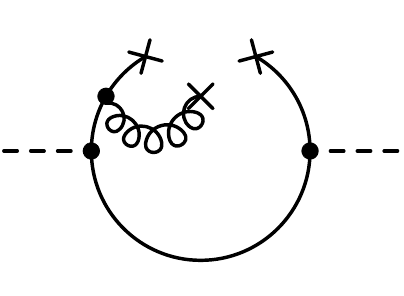}}}^{\!\!\!\!\!\!\text{(b)}\!\!}
							+ \parbox[c][15mm][c]{20mm}{\mbox{\includegraphics[width=2cm]{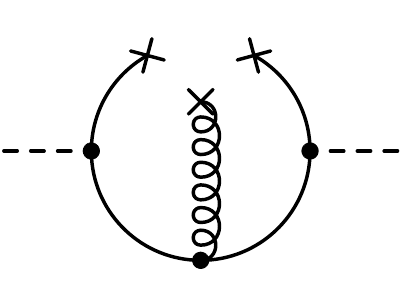}}}^{\!\!\!\!\!\!\text{(c)}\!\!}
							+ \parbox[c][15mm][c]{20mm}{\mbox{\includegraphics[width=2cm]{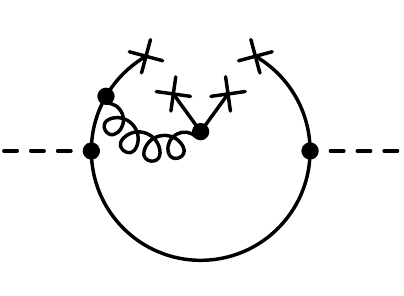}}}^{\!\!\!\!\!\!\text{(d)}\!\!}
							+ \parbox[c][15mm][c]{20mm}{\mbox{\includegraphics[width=2cm]{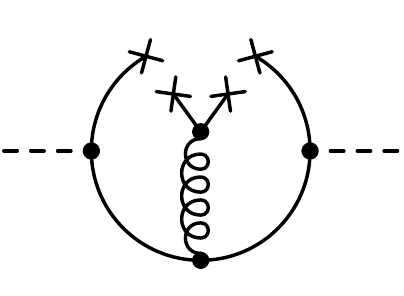}}}^{\!\!\!\!\!\!\text{(e)}\!\!}
							+	\ldots	\nonumber
\end{align*}
\egroup
\caption{Diagrammatic representation of the contribution \eqref{eq:2qterms}, where a selection of topologically relevant diagrams are displayed as in Fig.~\ref{fig:diagramPiAlpha0(0)}, too.}%
\label{fig:diagramPiAlpha0(2)}%
\end{figure}

Employing the expansion \eqref{eq:pertqprop} in \eqref{eq:fulcontractedterms} generates a series of terms where a (few) soft gluon(s) couple to quark, gluon and quark-gluon condensates (cf.\ Fig.~\ref{fig:diagramPiAlpha0(0)}), while an analog series emerges from \eqref{eq:2qterms} utilizing the covariant expansion of the quark operator additionally, see Fig.~\ref{fig:diagramPiAlpha0(2)}.

Up to mass dimension 5 the corresponding infra-red stable Wilson coefficients can be found in \cite{hilger09,zsch11}, providing coefficients for the vacuum and medium-specific quark, gluon and mixed quark-gluon condensates listed below Eq.~\eqref{eq:Dto5}. These refer to diagrams (a$-$c) in Fig.~\ref{fig:diagramPiAlpha0(0)} and (a$-$c) in Fig.~\ref{fig:diagramPiAlpha0(2)} as well as diagrams associated to non-local condensation. Using the formulae in \cite{zsch11,hilger11} one also obtains Wilson coefficients of light four-quark condensates, where the corresponding tree-level diagrams (cf.\ Fig.~\ref{fig:diagramPiAlpha0(2)}, diagrams (d) and (e)) contain one soft-momentum gluon line.

Since we truncate here the series expansion of \eqref{eq:fulcontractedterms} and \eqref{eq:2qterms} at order $g^2$, light four-quark condensate contributions arise only from \eqref{eq:2qterms}, i.\,e.\ diagrams (d) and (e) in Fig.~\ref{fig:diagramPiAlpha0(2)}. The other diagrams in the upper line of Fig.~\ref{fig:diagramPiAlpha0(2)} refer to the two-quark (a) and the quark-gluon condensate contributions (b) and (c), respectively.

\section{Four-quark condensate contributions}
\label{sec:wilcoeffncond}

\subsection{Wilson coefficients}
\label{subsec:wilcoeff}

\begin{figure}%
\bgroup
\makeatletter\@fleqnfalse\makeatother
\begin{align}
	\Pi^\text{NLO}_\text{dim6}	& = \parbox[c][15mm][c]{20mm}{\mbox{\includegraphics[width=2cm]{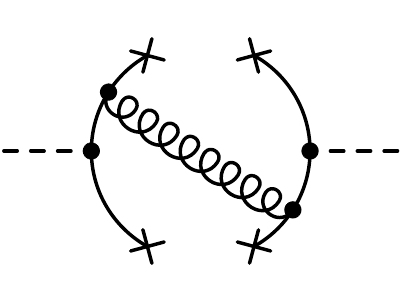}}}^{\!\!\!\!\!\!\text{(a)}\!\!}
																+ \parbox[c][15mm][c]{20mm}{\mbox{\includegraphics[width=2cm]{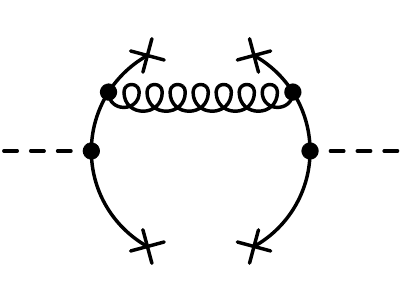}}}^{\!\!\!\!\!\!\text{(b)}\!\!} \nonumber
\end{align}
\egroup
\caption{Diagrammatic representation of hard-gluon contributions giving tree-level four-quark condensate terms for the correlator in next-to-leading order (NLO) perturbation theory $\alpha_s^1$.}%
\label{fig:Pi_4q_NLO}%
\end{figure}

We focus now on the evaluation of the light-quark condensate contributions in mass dimension 6 containing the particularly interesting four-quark condensates.\footnote{In medium, four-quark condensates can not be considered solely, but inclusion of corresponding light-quark condensates in mass dimension 6, which can not be reduced to four-quark condensates, is required in order to ensure a continuous transition to the vacuum (cf.\ the Appendix \ref{sec:calc4Q} for technical details).} For a handy notation, we denote by $\Pi_\text{dim6}$ those contributions to $\Pi^{(2)}$, cf.\ \eqref{eq:OPEs}, where dimension-6 light-quark condensates are involved (e.\,g.\ the four-quark condensate contributions in diagrams (d) and (e) in Fig.~\ref{fig:diagramPiAlpha0(2)}). We note that the gluon in the contributions to $\Pi^{(2)}$ may have an arbitrarily small momentum, and thus they are dubbed soft-gluon contributions. However, also tree-level four-quark contributions exist where the gluon must carry the full momentum $p$ of the meson (cf.\ Fig.~\ref{fig:Pi_4q_NLO}). These contributions are dubbed hard-gluon ones, and they arise with the next-to-leading order interaction term inserted into the correlator \cite{svz79,narison,pt84}.
However, their corresponding condensates contain heavy-quark operators and are neglected according to arguments in \cite{rry,narison05}: heavy quarks do hardly condense.
Note that we disregard higher-order light four-quark condensate contributions $\propto g^{2n}$ with $n \geq 2$, such as diagram (h) in Fig.~\ref{fig:diagramPiAlpha0(0)}.

In the light chiral limit, $m_q \rightarrow 0$, the exact result reads
\begin{align}
	\Pi_\text{dim6} (p) & = \frac{1}{3} \frac{1}{(p^2 - m_Q^2)^2} \left( 1 + \frac{1}{2} \frac{m_Q^2}{p^2 - m_Q^2} - \frac{1}{2} \frac{m_Q^4}{(p^2 - m_Q^2)^2} \right) g^2 \langle : \! O_1 \! : \rangle \nonumber\\[1mm]
	& + \frac{1}{9} \frac{1}{(p^2 - m_Q^2)^3} \left( p^2 - 4 \frac{(vp)^2}{v^2} \right) \Big[ \langle : \! g^2 O_1 - \frac{2}{v^2} \left( g^2 O_2 - 2 g O_3 + 6g O_4 \right) \! : \rangle \nonumber\\
	& \qquad\qquad - \langle : \! g^2 O_1 - \frac{2}{v^2} \left( g^2 O_2 + g O_5 \right) \! : \rangle  + \frac{2}{v^2} \langle : \!  g^2 O_2 + 3g O_5 - g O_6  \! : \rangle \nonumber\\
	& \qquad\qquad + \frac{2}{v^2} g \langle : \! O_3 \! : \rangle - \frac{3}{2}\langle : \! 3 g^2 O_1 - \frac{4}{v^2} \left( g^2 O_2 + 2g O_4 - g O_7 \right) \! : \rangle \Big] \nonumber\\[1mm]
	& - \frac{2}{15} \frac{1}{(p^2 - m_Q^2)^4} \left( p^4 - 7 p^2 \frac{(vp)^2}{v^2} + 6 \frac{(vp)^4}{v^4} \right) \Big[ \frac{2}{v^2} g \langle : \! O_3 \! : \rangle \nonumber\\
	& \qquad\qquad + \langle : \! g^2 O_1 - \frac{2}{v^2} \left( g^2 O_2 - 2 g O_3 + 6g O_4 \right) \! : \rangle \Big] \nonumber\\[1mm]
	& + \frac{1}{30} \frac{1}{(p^2 - m_Q^2)^4} \left( p^4 - 12 p^2 \frac{(vp)^2}{v^2} + 16 \frac{(vp)^4}{v^4} \right) \langle : \! g^2 O_1 - \frac{48}{v^4} O_8 \! : \rangle \nonumber\\[1mm]
	& - 2 \frac{m_Q}{(p^2 - m_Q^2)^3} \frac{(vp)}{v^2} \Big[ g^2 \langle : \! O_9 \! : \rangle + 2 g \left( \langle : \! O_{10} \! : \rangle + \langle : \! O_{11} \! : \rangle \right) \nonumber\\
	& \qquad\qquad - \frac{1}{3} g \langle : \! O_{12} \! : \rangle + \frac{1}{3} g \langle : \! O_{13} \! : \rangle \Big] \nonumber\\[1mm]
	& + \frac{8}{3} \frac{m_Q}{(p^2 - m_Q^2)^4} \frac{(vp)}{v^2} \left( p^2 - \frac{(vp)^2}{v^2} \right) \Big[ g^2 \langle : \! O_9 \! : \rangle + \frac{3}{2} g \langle : \! O_{10} \! : \rangle \Big] \nonumber\\[1mm]
	& - 8 \frac{m_Q}{(p^2 - m_Q^2)^4} \frac{(vp)}{v^4} \left( p^2 - 2 \frac{(vp)^2}{v^2} \right) \langle : \! O_{14} \! : \rangle \, ,
	\label{eq:result1P}
\end{align}
where the operators $O_k$ are listed in Tab.~\ref{tab:condoperators}.
\renewcommand{\arraystretch}{1.5}
\begin{table}%
\centering
\caption[small,bf]{List of light-quark operators of mass dimension 6 forming condensates which enter the results for soft-gluon diagram contributions to $\Pi^{(2)}$ (cf.\ Fig.~\ref{fig:diagramPiAlpha0(2)}) emerging from \eqref{eq:2qterms}. $\sum_f$ means summing over light quark flavours $u,d (,s)$. The quantity $\slashed{v}$ denotes the medium four-velocity contracted with a Dirac matrix and $(vD)$ its contraction with the covariant derivative.
}
\begin{tabular}{c p{0.1cm} l p{1cm} r p{0.1cm} l p{1cm} r p{0.1cm} l}
	\hline
	\multicolumn{1}{c}{$k$} && $\qquad O_k $	&&	\multicolumn{1}{c}{$k$} && $\qquad O_k $	&&	\multicolumn{1}{c}{$k$} && $\qquad O_k $	\\
	\cline{1-1}\cline{3-3}\cline{5-5}\cline{7-7}\cline{9-9}\cline{11-11}
	1				&&	$	\displaystyle \bar{q} \gamma^\nu t^A q \sum_f{ \bar{q}_f \gamma_\nu t^A q_f}  $	&&
	6				&&	$ \displaystyle \bar{q} \slashed{v} \sigma^{\nu\mu} \left[ (v i D) , G_{\nu\mu} \right] q $ &&
	11			&&	$ \displaystyle \bar{q} (v i \overset{{}_\leftarrow}{D}) \sigma G q $ \\
	2				&&	$	\displaystyle \bar{q} \slashed{v} t^A q \sum_f{ \bar{q}_f \slashed{v} t^A q_f} $	&&
	7				&&	$ \displaystyle \bar{q} i \overset{ {}_\leftarrow}{D}_\mu \gamma_5 \slashed{v} G_{\nu\lambda} q \varepsilon^{\mu\nu\lambda\tau} v_\tau $	&&
	12			&&	$ \displaystyle \bar{q} \sigma^{\mu\nu} \left[ i D_\mu , G_{\nu\lambda} \right] v^\lambda q $ \\
	3				&&	$ \displaystyle \bar{q} (v i \overset{{}_\leftarrow}{D}) \sigma G \slashed{v} q $	&&
	8				&&	$ \displaystyle \bar{q} (v i \overset{{}_\leftarrow}{D})^3 \slashed{v} q $	&&
	13			&&	$ \displaystyle \bar{q} \sigma^{\nu\mu} \left[ (v i D) , G_{\nu\mu} \right] q $ \\
	4				&&	$ \displaystyle \bar{q} (v\overset{{}_\leftarrow}{D}) \gamma^\mu G_{\nu\mu} v^\nu q $	&&
	9				&&	$	\displaystyle \bar{q} t^A q \sum_f{ \bar{q}_f \slashed{v} t^A q_f}  $ &&
	14			&&	$ \displaystyle \bar{q} (v i \overset{{}_\leftarrow}{D})^3 q $ \\
	5				&&	$ \displaystyle \bar{q} \gamma^\nu \left[ (vD) , G_{\nu\mu} \right] v^\mu q $	&&
	10			&&	$ \displaystyle \bar{q} \sigma G (v i D) q $ &&
					&&	\\
	\hline
\end{tabular}
\label{tab:condoperators}
\end{table}
The incorporated four-quark operators are obtained from the perturbative quark propagator exploiting the gluonic equation of motion. Even in medium, the number of such operators is limited, because the equation of motion predetermines Dirac and color structures. Current-current correlators with a single quark flavour $q$ form three different four-quark operators (of this origin) according to Eq.~\eqref{eq:conds} which are invariant under parity and time-reversal transformations. The corresponding condensates, i.\,e.\ $\langle O_k \rangle$ with $k=1$, 2 and 9, are classified as full condensates with indices 2v, 2v$'$ and 2vs in table~1 (for $q=q_f$) and with indices 4v, 4v$'$, 4vs and 6vs in table~2 (for $q\neq q_f$) of Ref.~\cite{thomas07} which provides an exhaustive list of independent light four-quark condensates.

Only the first line in \eqref{eq:result1P} contains the vacuum contribution whereas the remaining terms are medium-specific and consequently must vanish in vacuum \cite{buchheim14QCD14}.
Furthermore, it has been shown by consistency arguments alone that the particular linear combinations of operators collected in Tab.~\ref{tab:MedSpecOpCombinations} must vanish identically in vacuum. This imposes vacuum constraints as interrelations among the operators of Tab.~\ref{tab:condoperators}, in particular also between terms which already occur in vacuum, i.\,e.\ $O_1$, and those which additionally and exclusively enter in the medium. Note that vacuum specific terms additionally exhibit an own medium dependence.
\renewcommand{\arraystretch}{1.5}
\begin{table}%
\centering
\caption[small,bf]{List of medium-specific light-quark operator combinations in mass dimension 6 incorporating operators related to $O_1$ which already occurs in vacuum.
}
\begin{tabular}{l}
	\hline
	$ g^2 O_1 - \frac{2}{v^2} \left( g^2 O_2 - 2 g O_3 + 6g O_4 \right) $ \\
	$ g^2 O_1 - \frac{2}{v^2} \left( g^2 O_2 + g O_5 \right) $ \\
	$ \frac{2}{v^2} \left(  g^2 O_2 + 3g O_5 - g O_6 \right) $ \\
	$ \frac{2}{v^2} g O_3 $ \\
	$ 3 g^2 O_1 - \frac{4}{v^2} \left( g^2 O_2 + 2g O_4 - g O_7 \right) $ \\
	$ g^2 O_1 - \frac{2}{v^2} \left( g^2 O_2 - 2 g O_3 + 6g O_4 \right) $ \\
	$ g^2 O_1 - \frac{48}{v^4} O_8 $ \\
	\hline
\end{tabular}
\label{tab:MedSpecOpCombinations}
\end{table}

In order to test our computational procedure we consider the light four-quark condensate contributions of pseudo-scalar $D$ mesons in vacuum. Employing the Borel transformation and after factorization of the four-quark condensates (cf.\ Section \ref{sec:HQEandFac}), we recover the result first calculated by Aliev and Eletsky \cite{alieveletsky} and confirmed by Narison~\cite{narison01}
\begin{align}
	\widehat{\Pi}_\text{4q}^\text{vac} (M) = - \frac{16\pi}{27} \frac{e^{-m_Q^2/M^2}}{M^2} \bigg\{ 1 - \frac{1}{4} \frac{m_Q^2}{M^2} - \frac{1}{12}\frac{m_Q^ 4}{M^4} \bigg\} \alpha_\text{s} \kappa_1 \langle \bar{q}q \rangle_0^2 \, .
	\label{eq:resultNarison}
\end{align}
The result of Hayashigaki and Terasaki \cite{hay04} differs by a factor of $-2$ in the second of the three terms in $\{\ldots\}$ forming the Wilson coefficient. It is conceivable that they missed one of the three terms of mass dimension 6 leading to light four-quark condensate contributions eventually (cf.\ \cite{pt84} and the details in the Appendix \ref{sec:calc4Q}). In fact, omitting the four-quark term in \eqref{eq:missedByHaya} recovers the result in \cite{hay04}.

Having accomplished the evaluation of the OPE for light-quark condensates in mass dimension 6, one has to note that, in our expansion scheme, further diagrams contribute in leading order. These are, for example, the gluon condensates depicted in Fig.~\ref{fig:diagramPiAlpha0(0)} diagrams (d), (e) and (f) which deserve separate elaborations beyond the scope of this paper.

\subsection{Condensates and chiral transformations}

As stressed in the introduction, the chiral condensate $\langle \bar{q}q \rangle$ may serve as an order parameter of spontaneous chiral symmetry breaking (or can constitute an element thereof), since it is not invariant under chiral transformations
\begin{align}
	q_{\text{L},\text{R}} \longrightarrow q'_{\text{L},\text{R}} = e^{-i\Theta^a_{{}^{\text{L},\text{R}}}\tau^a} q_{\text{L},\text{R}} \, ,
	\label{eq:chiTrafo}
\end{align}
where $q = q_\text{L} + q_\text{R}$ is an $N_\text{f}$ dimensional light-flavour vector with $q_{\text{L},\text{R}} = \text{P}_{\text{L},\text{R}} q$ and $\text{P}_{\text{L},\text{R}} = (1 \mp \gamma_5)/2$. The matrices $\tau^a$ are the generators of the $\text{SU}(N_\text{f})_{\text{L},\text{R}}$ symmetry groups. Other quark condensates reveal invariant as well as non-invariant behavior under chiral transformations, thus, they are dubbed chirally even or chirally odd condensates, respectively. Four-quark condensates entering the OPE \eqref{eq:result1P} are of both kinds. Since these condensates (cf.\ Tab.~\ref{tab:condoperators}) are flavour singlet structures, such four-quark condensates containing exclusively $\gamma_\mu$ and/or $\gamma_5\gamma_\mu$ as Dirac structures are invariant under chiral transformations \eqref{eq:chiTrafo}.
Therefore, the first two entries in Tab.~\ref{tab:condoperators}, i.\,e.\ $\langle : \! O_1 \! : \rangle$ and $\langle : \! O_2 \! : \rangle$, are invariant under chiral transformations \eqref{eq:chiTrafo} whereas the condensate
\begin{align}
	\langle : \! O_9 \! : \rangle = \langle : \! \bar{q}_\text{R} t^A q_\text{L} \bar{q}_\text{L} \slashed{v} t^A q_\text{L} \! : \rangle + \langle : \! \bar{q}_\text{L} t^A q_\text{R} \bar{q}_\text{L} \slashed{v} t^A q_\text{L} \! : \rangle + (\text{L} \longleftrightarrow \text{R})
\end{align}
is chirally odd, i.\,e.\ it is turned into its negative for $|\Theta^a_\text{R} - \Theta^a_\text{L}| = (2k+1)\pi$ with integer~$k$. We note that this chirally odd four-quark condensate is medium-specific contrary to the chiral condensate. The chirally odd nature of $\langle : \! O_9 \! : \rangle$ can be also deduced from the difference of chiral partner spectra (Weinberg-type sum rules), where the dependence on chirally symmetric condensates cancels out \cite{hilger11}. Furthermore, the operator $O_9$ emerges from a commutator with the generator of the axial-vector transformation \cite{thomasdis} similarly to operators providing potential order parameters, e.\,g.\ the chiral condensate. Thus, the chirally odd four-quark condensate contributions may give insight to the breaking patterns of chiral symmetry as well as symmetry restoration scenarios \cite{hohlerrapp12,holthohlerrapp13,hohlerrapp14}.

\section{Estimates of four-quark condensate contributions}
\label{sec:HQEandFac}

Once the evaluation of the OPE is completed, even in a truncated form and according to a particular organization of the nested multiple expansion schemes, one needs numerical values of the various condensates. The low-mass dimension condensates are constrained fairly well, even with some debate on the gluon condensates \cite{shuryakbook}.\footnote{We also refer the interested reader to \cite{brodsky12} for a general discussion of condensates and their relation to hadron wave functions with emphasis on the light-front formulation.} The mass dimension-6 four-quark condensates are less well investigated. They enter QCD sum rules in different combinations, as exemplified, for instance, in \cite{thomas07} for the nucleon and in \cite{hilger12} for the $\rho$ meson.

To arrive at some numerical estimates of the impact of the light four-quark condensates we employ tentatively the factorization hypothesis, being aware of its limited reliability and lacking foundation \cite{leupold05,koike}. Despite the validity of the factorization ansatz for an infinite number of colors, its accuracy for QCD is still questionable. Factorization of four-quark condensates is based on the ground state saturation hypothesis. Accordingly, only the ground state is assumed to yield a relevant contribution after insertion of a complete set of hadronic eigen states into the four-quark condensate. In \cite{svz79} the contribution of the lightest hadronic state, the pion state, is estimated as $1/20$ of the ground state contribution. Thus, the four-quark condensate is assumed to factorize into two ground state expectation values of two-quark operators. In a medium, two different two-quark condensates exist, \mbox{$\langle : \! \bar{q}q \! : \rangle$} and \mbox{$\langle : \! \bar{q}\gamma^\mu q \! : \rangle$}, where the latter one is employed as \mbox{$\langle : \! \bar{q}\slashed{v}q \! : \rangle v^\mu/v^2$} after projection of the Lorentz index. Factorization formulae for in-medium contributions can be found in \cite{jin}. Our investigation uses
\begin{align}
	& \langle : \! \bar{q} \Gamma_1 t^A q \bar{q} \Gamma_2 t^A q \! : \rangle = - \frac{\kappa(\Gamma_1,\Gamma_2)}{36} \Big\{ \langle : \! \bar{q}q \! : \rangle^2 \text{Tr}_\text{D} \left[ \Gamma_1 \Gamma_2 \right] \nonumber\\
	& \phantom{\langle : \! \bar{q} \Gamma_1 t^A q \bar{q} \Gamma_2 t^A q \! : \rangle = - \frac{1}{36}} + \langle : \! \bar{q} q \! : \rangle \langle : \! \bar{q} \gamma_\mu q \! : \rangle \left( \text{Tr}_\text{D} \left[ \Gamma_1 \gamma^\mu \Gamma_2 \right]  + \text{Tr}_\text{D} \left[ \gamma^\mu\Gamma_1 \Gamma_2 \right] \right) \nonumber\\
	& \phantom{\langle : \! \bar{q} \Gamma_1 t^A q \bar{q} \Gamma_2 t^A q \! : \rangle = - \frac{1}{36}} + \langle : \! \bar{q} \gamma_\mu q \! : \rangle \langle : \! \bar{q} \gamma_\nu q \! : \rangle \text{Tr}_\text{D} \left[ \gamma^\mu\Gamma_1 \gamma^\nu \Gamma_2 \right] \Big\} \, ,
	\label{eq:fac1}	\\
	& \langle : \! \bar{q} \Gamma_1 t^A q \bar{q}_f \Gamma_2 t^A q_f \! : \rangle = 0 \, ,
	\label{eq:fac2}
\end{align}
where $\Gamma_1$ and $\Gamma_2$ denote the Dirac structures of the condensates and $q \neq q_f$ in Eq.~\eqref{eq:fac2}. The factors $\kappa(\Gamma_1,\Gamma_2)$ may be introduced to account for deviations from strict factorization, which is recovered for $\kappa(\Gamma_1,\Gamma_2) = 1$. The relevant expressions are listed in Tab.~\ref{tab:4qfac}.
\renewcommand{\arraystretch}{1.5}
\begin{table}%
\centering
\caption[small,bf]{List of light four-quark condensates $ \langle : \! O_k \! : \rangle$ entering the result \eqref{eq:result1P} in genuine and factorized form. Resulting from the two-quark condensates in linear density approximation, the fourth column depicts the density dependent and factorized four-quark condensates, utilized in Sec.~\ref{sec:numeval}.}
\begin{tabular}{rp{0.0cm}lp{0.1cm}lp{0.1cm}l
}
	\hline
	$ k $
			&&$ \langle : \! O_k \! : \rangle $
			&&	factorized
			&& density dependent\\
	\cline{1-1}\cline{3-3}\cline{5-5}\cline{7-7}
	1
			&&	$	\displaystyle \langle : \! \bar{q} \gamma^\nu t^A q \sum_f{ \bar{q}_f \gamma_\nu t^A q_f} \! : \rangle $
			&&	$ \displaystyle - \frac{2}{9} \kappa_1 \!\Big[ 2 \langle : \! \bar{q} q \! : \rangle^2 - \langle : \! \bar{q} \slashed{v} q \! : \rangle^2 / v^2 \Big] $
			&&	$ \displaystyle -\frac{4}{9} \kappa_1 \!\bigg[ \langle : \! \bar{q}q \! : \rangle_0^2 \Big( 1 - \frac{\sigma_N}{m_\pi^2 f_\pi^2} n \Big)^{\! 2} - \frac{9}{8} n^2 \bigg] $	\\
	2
			&&	$	\displaystyle \langle : \! \bar{q} \slashed{v} t^A q \sum_f{ \bar{q}_f \slashed{v} t^A q_f} \! : \rangle $
			&&	$ \displaystyle - \frac{1}{9} \kappa_2 \!\Big[ v^2 \langle : \! \bar{q} q \! : \rangle^2 + \langle : \! \bar{q} \slashed{v} q \! : \rangle^2 \Big] $
			&&	$ \displaystyle -\frac{1}{9} \kappa_2 \!\bigg[ \langle : \! \bar{q}q \! : \rangle_0^2 \Big( 1 - \frac{\sigma_N}{m_\pi^2 f_\pi^2} n \Big)^{\! 2} + \frac{9}{4} n^2 \bigg] $	\\
	9
			&&	$	\displaystyle \langle : \! \bar{q} t^A q \sum_f{ \bar{q}_f \slashed{v} t^A q_f} \! : \rangle $
			&&	$	\displaystyle - \frac{2}{9} \kappa_3 \langle : \! \bar{q} q \! : \rangle \langle : \! \bar{q} \slashed{v} q \! : \rangle $
			&&	$	\displaystyle - \frac{1}{3} \kappa_3 \langle : \! \bar{q}q \! : \rangle_0 \Big( 1 - \frac{\sigma_N}{m_\pi^2 f_\pi^2} n \Big) n $\\
	\hline
\end{tabular}
\label{tab:4qfac}
\end{table}

However, Tab.~\ref{tab:4qfac} exhibits further issues which arise due to factorization of four-quark condensates. Condensates originally considered chirally even, such as \mbox{$\langle : \! O_1 \! : \rangle$} and \mbox{$\langle : \! O_2 \! : \rangle$}, factorize into powers of the chiral condensate \mbox{$\langle : \! \bar{q}q \! : \rangle$} which is genuinely chirally odd. Since factorization changes the behavior of four-quark condensates under chiral transformations, the transformation properties of the OPE and, therefore, of the (operator product) expanded correlator are altered. A procedure analog to the one in \cite{hilger12} can overcome such artifacts.

\section{Numerical evaluation}
\label{sec:numeval}

The Borel transform \cite{cohen95,furnstahl} of the light-quark condensate contributions in the meson rest frame, $p_\mu=(p_0,\vec{0})$, $v_\mu=(1,\vec{0})$, and after a Wick rotation $p_0 = i \omega$, reads
\begin{subequations}
\label{eq:resultsBorel}
\begin{align}
	\widehat{\Pi}^\text{even}_\text{dim6} (M^2) & = \frac{1}{3} \frac{e^{-m_Q^2/M^2}}{M^2} \left( 1 - \frac{1}{4} \frac{m_Q^2}{M^2} - \frac{1}{12} \frac{m_Q^4}{M^4} \right) g^2 \langle : \!O_1 \! : \rangle \nonumber\\[1mm]
	& - \frac{1}{3} \frac{e^{-m_Q^2/M^2}}{M^2} \left( 1 - \frac{1}{2} \frac{m_Q^2}{M^2} \right) \Big[ \langle : \! g^2 O_1 - 2 \left( g^2 O_2 - 2 g O_3 + 6g O_4 \right) \! : \rangle \nonumber\\
	& \qquad\qquad - \langle : \! g^2 O_1 - 2 \left( g^2 O_2 + g O_5 \right) \! : \rangle + 2 \langle : \! g^2 O_2 + 3g O_5 - g O_6  \! : \rangle \nonumber\\
	& \qquad\qquad + 2 g \langle : \! O_3 \! : \rangle - \frac{3}{2} \langle : \! 3 g^2 O_1 - 4 \left( g^2 O_2 + 2g O_4 - g O_7 \right) \! : \rangle \Big] \nonumber\\[1mm]
	& + \frac{1}{6} \frac{e^{-m_Q^2/M^2}}{M^2} \left( 1 - \frac{m_Q^2}{M^2} + \frac{1}{6} \frac{m_Q^4}{M^4} \right) \langle : \! g^2 O_1 - 48 O_8 \! : \rangle	\, ,
	\label{eq:even}\\[3mm]
	\widehat{\Pi}^\text{odd}_\text{dim6} (M^2) & = \frac{e^{-m_Q^2/M^2}m_Q}{M^4} \Big[ g^2 \langle : \! O_9 \! : \rangle + 2 g \left( \langle : \! O_{10} \! : \rangle + \langle : \! O_{11} \! : \rangle \right) \nonumber\\
	& \qquad\qquad - \frac{1}{3} g \langle : \! O_{12} \! : \rangle + \frac{1}{3} g \langle : \! O_{13} \! : \rangle \Big] \nonumber\\[1mm]
	& - 4 \frac{e^{-m_Q^2/M^2}m_Q}{M^4} \left( 1 - \frac{1}{3} \frac{m_Q^2}{M^2} \right) \langle : \! O_{14} \! : \rangle \, ,
	\label{eq:odd}
\end{align}
\end{subequations}
where even and odd part w.\,r.\,t.\ $p_0$ have been separated.
As described in Subsec.~\ref{subsec:wilcoeff} the medium-specific condensates entering the results \eqref{eq:resultsBorel} must vanish for zero nuclear density per construction. This can be satisfied for the odd OPE~\eqref{eq:odd} by vanishing of the square bracketed terms and $\langle : \! O_{14} \!: \rangle$, and it implies non-trivial constraints on the vacuum expectation values of operators in the even OPE~\eqref{eq:even} containing the medium velocity $v_\mu$. These vacuum constraints can be formulated in terms of a system of linear equations (for the entries listed in Tab.~\ref{tab:MedSpecOpCombinations}) which can be shown to be solvable and ensures vanishing of all medium-specific condensates in vacuum.

As, however, the density dependence of the medium-specific light-quark condensates in mass dimension 6 is still unrestricted, we assume that it is dominated by the four-quark condensate contribution whose medium dependence can be estimated by factorization (cf.\ Tab.~\ref{tab:4qfac}), i.\,e.\ only $O_i$ with $i = 1$, 2 and 9 exhibit a medium dependence. The other condensates are constant w.\,r.\,t.\ density and temperature and remain at their vacuum values dictated by vacuum constraints. This reduces the numerical values of the medium-specific condensates to the density dependent terms of the four-quark condensate contributions of the fourth column in Tab.~\ref{tab:4qfac}.
Furthermore, the factorization parameters $\kappa_i$ are also related via these constraints. One has
\begin{align}
	\kappa_2 (n=0) = 3 \kappa_1 (n=0) \, .
\end{align}
In order to indicate the numerical evaluation of $\widehat{\Pi}_\text{dim6}$ under these assumptions we introduce the notation $\widehat{\Pi}_\text{4q-dom}$.

Note that we utilize four-quark condensates beyond linear density approximation with the following reasoning. Taking into account only the linear density terms in the fourth column of Tab.~\ref{tab:4qfac} yields a chirally odd condensate, namely $\langle: \!  O_9 \! : \rangle$, which does not vanish at higher densities, i.\,e.\ exhibits no signals of chiral restoration, in contradiction with the chiral condensate. However, employing the linear density approximation to the two-quark condensates entering the factorized four-quark condensates provides a quadratic density dependence which overcomes such issues. (It turns out that imposing the above described constraints and using the linear density dependence of factorized four-quark condensates, only the odd term carries a medium dependence, whereas the medium dependence of the even term -- including the vacuum specific term -- completely cancels out.)

Assuming constant $\kappa$'s w.\,r.\,t.\ density and using the notation $\langle O_i^n \rangle = \langle O_i \rangle - \langle O_i^0 \rangle$ with the vacuum part $\langle O_i^0 \rangle = - \frac{a_i}{9} \kappa_i \langle \bar{q}q \rangle_0^2$ for $i=1$ and $2$ with $a_1 = 4$ and $a_2 = 1$ the Borel transformed density dependent result reads
\begin{subequations}
\label{eq:results4qdom}
\begin{align}
	\widehat{\Pi}^\text{even}_\text{4q-dom} (M^2) & = \frac{1}{3} \frac{e^{-m_Q^2/M^2}}{M^2} \left( 1 - \frac{1}{4} \frac{m_Q^2}{M^2} - \frac{1}{12} \frac{m_Q^4}{M^4} \right) g^2 \langle O_1 \rangle \nonumber\\[1mm]
	& - \frac{1}{3} \frac{e^{-m_Q^2/M^2}}{M^2} \left( 1 - \frac{1}{2} \frac{m_Q^2}{M^2} \right) \Big[ 2 g^2 \langle O_2^n \rangle - \frac{3}{2} g^2 \langle 3 O_1^n - 4 O_2^n \rangle \Big] \nonumber\\[1mm]
	& + \frac{1}{6} \frac{e^{-m_Q^2/M^2}}{M^2} \left( 1 - \frac{m_Q^2}{M^2} + \frac{1}{6} \frac{m_Q^4}{M^4} \right) g^2 \langle O_1^n \rangle	\, ,
	\label{eq:even4qdom}\\[3mm]
	\widehat{\Pi}^\text{odd}_\text{4q-dom} (M^2) & = \frac{e^{-m_Q^2/M^2}m_Q}{M^4} g^2 \langle O_9 \rangle \, ,
	\label{eq:odd4qdom}
\end{align}
\end{subequations}
where condensates are displayed without normal ordering, since we introduce physical condensates by renormalization of the normal-ordered condensates to one-loop level as described in \cite{jaminmunz}. According to arguments in \cite{zsch11,jaminmunz} for gluon condensates we obtain no further terms upon renormalization since four-quark condensate contributions are already of order $\alpha_\text{s}$.
Note that the vacuum constraints are the minimal requirements for a consistent vacuum limit. Any more sophisticated medium dependence must go beyond factorization with constant parameters and/or also impose medium dependence to the terms $O_i$ with $i = 3, \ldots , 8$ and $10, \ldots , 14$ in Tab.~\ref{tab:condoperators}.

Our numerical evaluation employs the values of the condensates including their nucleon density dependencies presented in \cite{hilger09}. We resort here to the four-quark condensate factorization parameters $\kappa_1 = 1$, $\kappa_2 = 3$ and $\kappa_3 = 1$, bearing in mind that the actual values may considerably deviate from the assumed values. We use for the heavy-quark mass $m_Q = \unit{1.5}{GeV}$ and determine the strong coupling from the one loop result $\alpha_\text{s} 	= 4\pi/[ (11 - 2N_\text{f}/3) \log(\mu^2/\Lambda_\text{QCD}^2) ]$ with the renormalization scale $\mu = \unit{1}{GeV}$, the dimensional QCD parameter $\Lambda_\text{QCD} = \unit{0.25}{GeV}$ and the number of light-quark flavours $N_\text{f} = 3$. The nucleon saturation density is $n = \unit{0.15}{fm^{-3}}$.

Subject to the following investigation is the OPE side of the Borel transformed QCD sum rule of $qQ$ mesons (cf.\ Eq.~\eqref{eq:Dto5})
\begin{subequations}
\label{eq:Dto5+4q}
\begin{align}
	\widehat{\Pi}^\text{even}(M^2) 	&	= C_0 + e^{-m_Q^2/M^2} \sum_{k=1}^6 c^\text{even}_k(M^2) \, \langle \mathcal{O}_k \rangle^\text{even}
																		+ \widehat{\Pi}^\text{even}_\text{4q-dom}(M^2)	\, ,
	\label{eq:Deven+4qApp}\\
	\widehat{\Pi}^\text{odd}(M^2) 	&	= e^{-m_Q^2/M^2} \sum_{k=1}^3 c^\text{odd}_k(M^2) \, \langle \mathcal{O}_k \rangle^\text{odd}
																		+ \widehat{\Pi}^\text{odd}_\text{4q-dom}(M^2) \, .
	\label{eq:Dodd+4qApp}
\end{align}
\end{subequations}
A full sum rule analysis up to condensates of mass dimension 5 shows that the density dependence of the mass centroid for $D$ and $\overline{D}$ mesons is mainly determined by the even part, while the mass splitting of the meson anti-meson pair is influenced by the odd part of the OPE \cite{hilger09}.\footnote{If four-quark condensates are included in linear density approximation into the sum rule their medium-modification effects only the odd OPE, and thus only the mass splitting of $D$ and $\overline{D}$. The mass centroid remains unaffected by such four-quark condensate contributions.} To get an estimate on the impact of light-quark condensate contributions in mass dimension 6, especially four-quark condensate contributions, on the meson properties we compare them to contributions of condensates up to mass dimension 5.

\begin{figure}%
\centering
\includegraphics{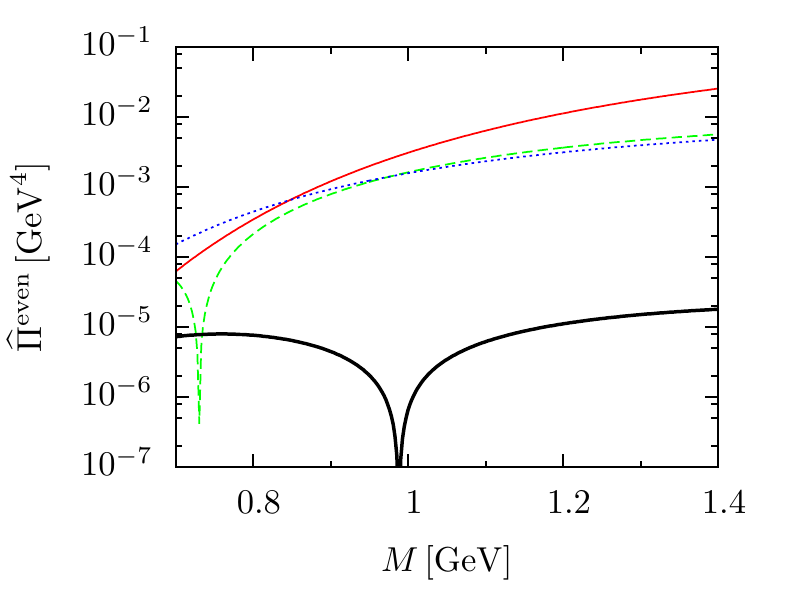}
\caption{Various contributions to the even OPE part of the $qQ$ sum rule at saturation density. The red solid curve depicts the perturbative term subtracted by the continuum contribution of the phenomenological side of the sum rule according to quark-hadron duality, i.\,e.\ $\overline{C}_0 = C_0 - \frac{1}{\pi} \left( \int_{-\infty}^{\omega_0^-} + \int_{\omega_0^+}^{+\infty} \right) d\omega\, \omega\, e^{-\omega^2/M^2} \text{Im}C_0(\omega^2) = \frac{1}{\pi}\int_{m_Q^2}^{\omega_0^2} ds\, e^{-s/M^2} \text{Im}C_0(s)
$, where we assume $\omega_0^+ = - \omega_0^- = \omega_0$ for the second equality with the symmetric continuum threshold parameter $\omega_0^2 = \unit{6}{GeV^2}$ (cf.~\cite{hilger09}). The green dashed curve is the modulus of the power correction $e^{-m_Q^2/M^2} \sum\limits_{k=1}^6 c^\text{even}_k \, \langle \mathcal{O}_k \rangle^\text{even}$ of the in-medium OPE \eqref{eq:Deven} up to mass dimension 5 according to \cite{hilger09}. The contribution of the chiral condensate, i.\,e.\ $-e^{-m_Q^2/M^2}\, m_Q \langle \bar{q}q \rangle$, is shown by the blue dotted curve. The modulus of the four-quark condensate contribution $\widehat{\Pi}^\text{even}_\text{4q-dom}$ \eqref{eq:even} is displayed by the thick solid black curve. For the curves with sign flips in the depicted region the left (right) branch of the green dashed (thick solid black) curve originates from negative values.}%
\label{fig:Comparison}%
\end{figure}
The QCD sum rules of $qQ$ mesons are governed by the perturbative and the chiral condensate contributions (cf.\ Fig.~\ref{fig:Comparison}). The purely perturbative contribution is even more prominent than displayed in Fig.~\ref{fig:Comparison}, where we already subtracted the continuum contribution of the spectral density of the sum rule employing the quark-hadron duality. The continuum threshold $\omega_0^2$, which needs to be chosen to meet stability criteria of the sum rule, is set to the typical value of $\unit{6}{GeV^2}$ for our investigation (cf.~\cite{narison05}). The chiral condensate contribution (blue dotted curve) has the strongest impact on the sum rule among the power corrections (green dashed curve). However, at typical Borel masses $M = 0.9 - \unit{1.3}{GeV}$ \cite{hilger09} further condensates contribute weakly.

\begin{figure}%
\includegraphics{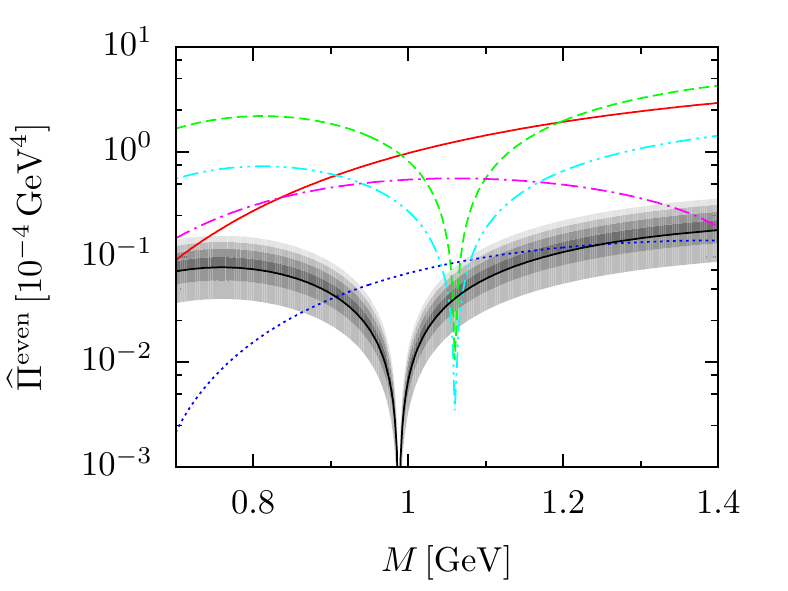}
\hfill
\includegraphics{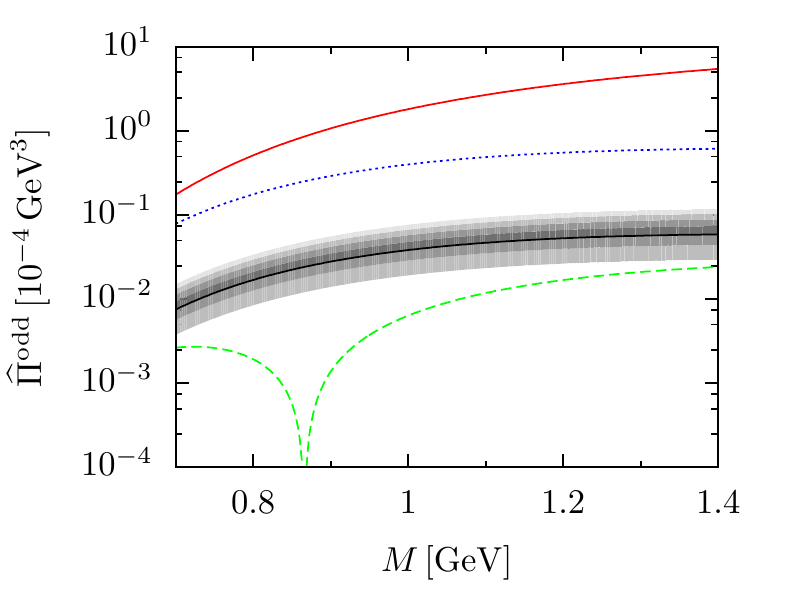}
\caption{Modulus of the individual contributions to the even (left panel) and the odd (right panel) OPE \eqref{eq:Dto5} at saturation density up to mass dimension 5 according to \cite{hilger09} supplemented by the four-quark condensate contributions exhibited by the solid black curves and the contours with $\kappa_1 = \kappa_2 / 3 = \kappa_3 \in [0.5,2]$. Using the notation of the contributions according to the condensates $\langle \mathcal{O}_k \rangle^\text{even,odd}$ listed below \eqref{eq:Dto5} for the even OPE on the left panel the following line code applies: $\langle \mathcal{O}_2 \rangle^\text{even}$ -- red solid curve, $\langle \mathcal{O}_3 \rangle^\text{even}$ -- green dashed curve, $\langle \mathcal{O}_4 \rangle^\text{even}$ -- blue dotted curve, $\langle \mathcal{O}_5 \rangle^\text{even}$ -- magenta dot-dashed curve and $\langle \mathcal{O}_6 \rangle^\text{even}$ -- cyan dot-dot-dashed curve. For the  curves with sign flips in the depicted region the left branches of the green dashed and cyan dot-dot-dashed curves and the right branch of the black solid curve originate from negative values. On the right panel the red solid curve depicts the $\langle \mathcal{O}_1 \rangle^\text{odd}$ contribution, the green dashed curve displays the sign flipped $\langle \mathcal{O}_2 \rangle^\text{odd}$ contribution and the blue dotted curve shows the $\langle \mathcal{O}_3 \rangle^\text{odd}$ contribution, where the right branch originates from negative values.}%
\label{fig:singleContrib_satDensity}%
\end{figure}
The absolute numerical values of light four-quark condensate contributions $\widehat{\Pi}^\text{even}_\text{4q-dom}$ (thick black solid curve) are two orders of magnitude below the contribution of the chiral condensate in the presumed Borel window, due to the heavy quark mass accompanying the chiral condensate acting as an amplification factor. Evaluation of individual contributions to the even and odd OPE exhibited in Fig.~\ref{fig:singleContrib_satDensity} shows the tendency of decreasing values of the contributions of the condensates with increasing mass dimension. The four-quark condensate contributions are of mass dimension 6 and therefore the highest order contribution of the evaluated OPE. Despite varying $\kappa_1 = \kappa_2 / 3 = \kappa_3$ between $0.5$ and 2 they are more than one order of magnitude smaller than most other contributions of the OPE up to mass dimension 5 yielding small absolute values which supports the convergence of the OPE. Moreover, trusting the order of magnitude of light four-quark condensate contributions exhibited in Fig.~\ref{fig:Comparison} lends credibility of the previous analyses, e.\,g.\ \cite{hilger09}, which were truncated at mass dimension 5.

Besides four-quark condensates also gluonic condensates contribute additionally to the in-medium OPE \eqref{eq:Dto5+4q} in mass dimension 6. Their contributions may numerically influence the OPE as strongly as the presented light-quark condensate terms, thus, they deserve separate further investigation. We choose the light four-quark condensates to serve as the starting point for the analysis of the OPE in mass dimension 6, because they are especially important in other meson sum rules as stressed in the introduction.

\section{Comparison of four-quark condensates in \texorpdfstring{$\boldsymbol \rho$}{rho} meson and \texorpdfstring{$\boldsymbol D$}{D} meson sum rules }
\label{sec:CompRhoD}

\begin{figure}%
\includegraphics[width=0.5\textwidth]{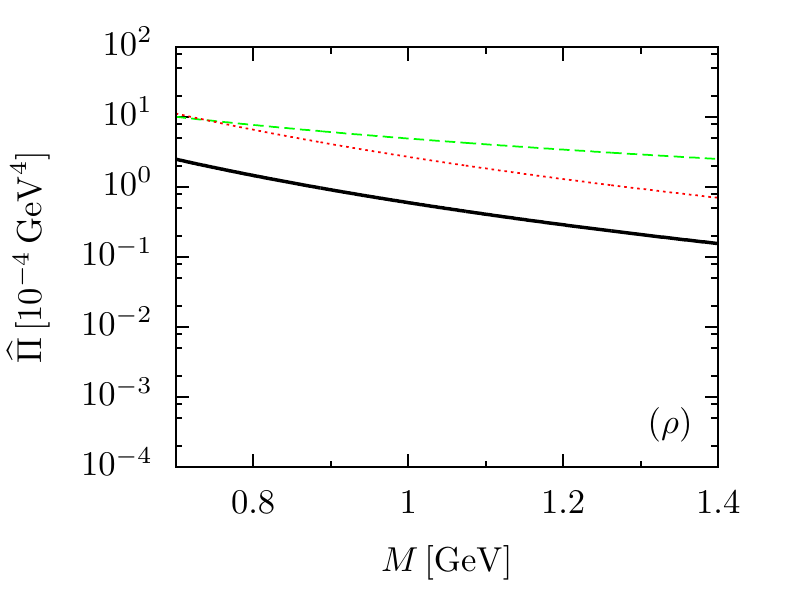}%
\hfill
\includegraphics[width=0.5\textwidth]{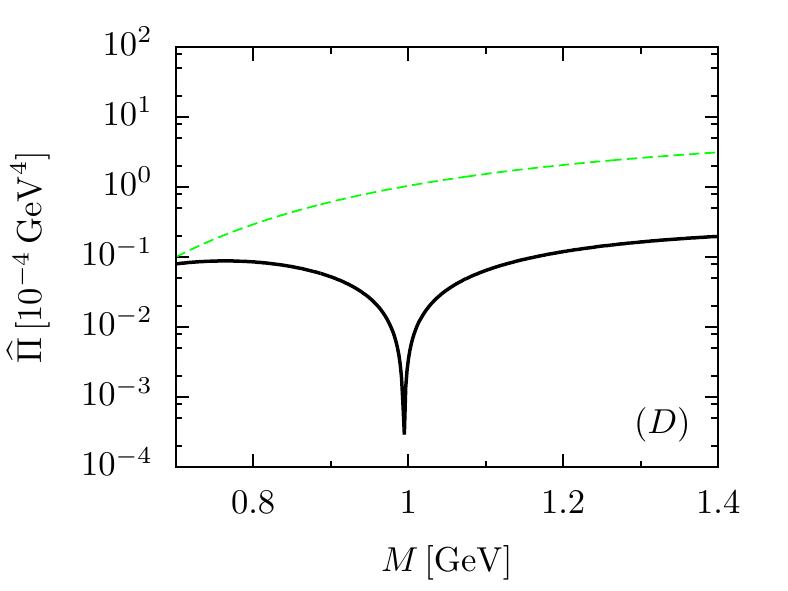} \\
\includegraphics[width=0.5\textwidth]{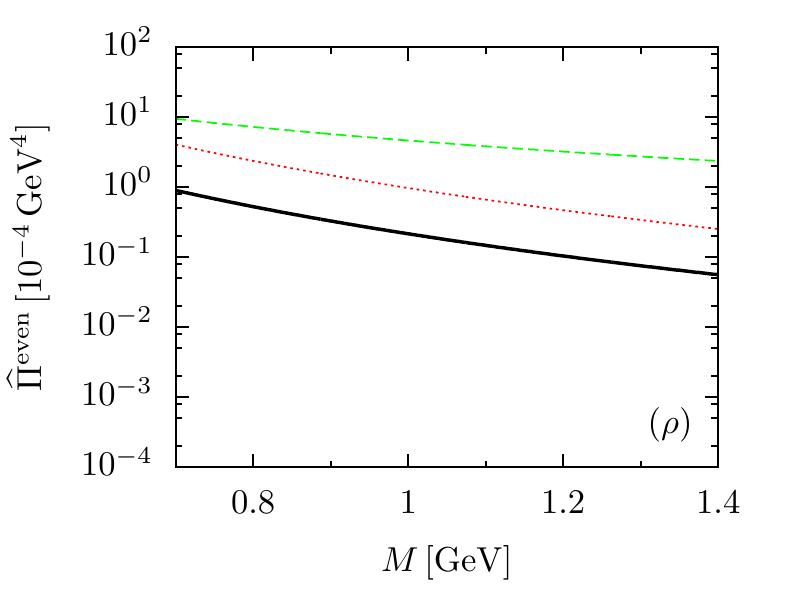}%
\hfill
\includegraphics[width=0.5\textwidth]{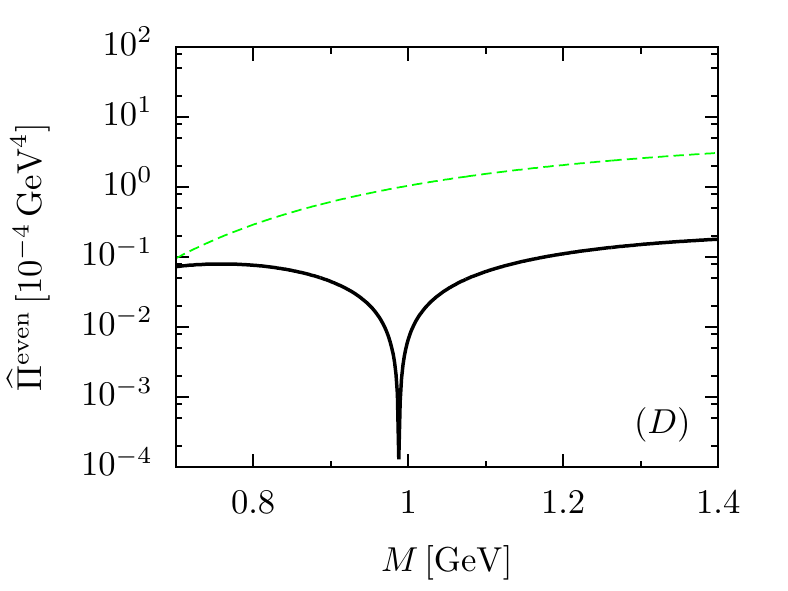}%
\caption{Comparison of $\rho$ meson (left panels) and $D$ meson (right panels) contributions in vacuum (upper panels) and at nucleon saturation density (lower panels). Dashed green curves denote the gluon condensate contributions and thick solid black curves display the modulus of light four-quark condensate contributions containing the soft gluon diagrams (d) and (e) in Fig.~\ref{fig:diagramPiAlpha0(2)}. The modulus of the four-quark condensate contribution of the $\rho$ meson from hard-gluon diagrams in Fig.~\ref{fig:Pi_4q_NLO} is depicted by the dotted red line.}%
\label{fig:DRho4qComp}%
\end{figure}

To compare with the $\rho$ meson OPE neglecting higher-twist terms, where the gluon and four-quark condensate contributions are of similar magnitude (cf.\ upper left panel in Fig.~\ref{fig:DRho4qComp}), we consider the vacuum four-quark condensate term of the $D$ meson (cf.\ upper right panel in Fig.~\ref{fig:DRho4qComp}). Its contribution is up to one order of magnitude smaller than the vacuum gluon condensate term. This order of magnitude difference arises from the $\rho$ meson OPE where the soft-gluon diagrams (d) and (e) in Fig.~\ref{fig:diagramPiAlpha0(2)} are supplemented by hard-gluon diagrams (a) and (b) in Fig.~\ref{fig:Pi_4q_NLO} whose numerical contributions exceed the soft-gluon contributions by a factor of five (cf.\ left panels in Fig.~\ref{fig:DRho4qComp}) and the gluon condensate term contributes half as much as the gluon condensate contribution in the $D$ meson sum rule. We argue that this can be disentangled as follows. While terms proportional to positive integer powers of the light-quark mass squared are neglected in the $\rho$ meson OPE, analogous heavy-quark terms significantly contribute to the $D$ meson OPE, where additionally the chiral condensate as well as the quark-gluon condensate are redefined to render the gluon condensate term free of infra-red mass singularities which is necessary in heavy-light systems \cite{jaminmunz}.

Wilson coefficients of the OPEs for $qQ$ meson systems exhibit non-monotonic behaviors for varying Borel mass parameters in contrast to light meson systems, due to the non-zero heavy-quark mass entering QCD sum rules of heavy-light systems as an additional scale. Non-negligible terms proportional to positive integer powers of the heavy-quark mass squared (cf.\ Eq.~\eqref{eq:resultNarison}) lead to roots of the Wilson coefficient in the Borel mass region near $M = \unit{1}{GeV}$ (cf.\ Fig.~\ref{fig:singleContrib_satDensity}), which is in the Borel window of $D$ meson sum rules resulting in small OPE contributions for terms with altering Wilson coefficients, such as light four-quark condensate contributions.

Studying the $\rho$ meson sum rule in the VOC (vanishing of chirally odd condensates) scenario (cf.\ Ref.~\cite{hilger12}), where the mass and/or width of the $\rho$ meson are evaluated in a hypothetical chirally symmetric world, the omission of medium-specific contributions\footnote{Medium-specific contributions of the $\rho$ meson OPE are usually denoted non-scalar or higher-twist terms~\cite{leupold98}.} is justified. In such a clear-cut scenario, the chirally odd objects, e.\,g.\ the chiral condensate, do not vanish due to their medium-modifications, but are set to zero, while the chirally symmetric condensates retain their vacuum values. In contrast, investigating signals of chiral restoration in the observed chirally broken world, chirally odd condensates are diminished due to an ambient medium. Thus, the inclusion of medium-specific contributions to the $\rho$ meson OPE is mandatory for a consistent in-medium description.
Four-quark condensate contributions without their medium-specific part exhibit an artificially strong medium dependence in comparison with the complete $D$ meson four-quark terms which show a minor medium dependence only (cf.\ Fig.~\ref{fig:DRho4qComp} lower panels compared to upper ones).
Furthermore, these medium-specific contributions also contain chirally odd objects, e.\,g.\ $\langle (\bar{\psi}\slashed{v}\gamma_5\lambda^a\tau_3\psi)^2 - (\bar{\psi}\slashed{v}\lambda^a\tau_3\psi)^2 \rangle$ traceable to the Gibbs averaged twist-4 operator $\langle ST (\bar{\psi}\gamma_\mu\gamma_5\lambda^a\tau_3\psi)(\bar{\psi}\gamma_\nu\gamma_5\lambda^a\tau_3\psi) \rangle$ (with adopted notation from \cite{hilger12}). However, identification of such chirally odd objects requires the decomposition of the non-scalar terms analogous to the procedure presented here for the $D$ meson and deserves further investigations.

\section{Summary}
\label{sec:disnconcl}

A systematic evaluation of the current-current correlator within the framework of QCD sum rules leads to a series of expectation values of combined QCD operators multiplied by Wilson coefficients. The seminal analysis of the $\rho$ meson \cite{hatsulee} points to the crucial impact of light four-quark operator structures (cf.\ \cite{hilger12,hohlerrapp12,holthohlerrapp13,hohlerrapp14}).
Driven by this insight we have evaluated the in-medium QCD sum rule for pseudo-scalar $\bar{q}Q$ and $\bar{Q}q$ mesons up to mass dimension 6 with emphasis on light four-quark condensate contributions, thus extending previous studies for vacuum \cite{shuryak82,alieveletsky,narison88,narison94,narison99,narison01,narison05,narison13,hay04} and medium \cite{hilger09,zsch11}.

Due to lacking information on precise numerical values of four-quark condensates, we employed tentatively the factorization hypothesis to estimate the numerical importance of light four-quark condensate contributions.  In contrast to the $\rho$ meson sum rule, the power corrections of higher mass dimension are obviously consecutively smaller, as mentioned already in \cite{rry} for vacuum and highlighted in \cite{hilger09} for in-medium situations. The heavy quark mass in the combination $m_Q \langle \bar{q}q \rangle$ provides a numerically large contribution to the OPE making it dominating. Having now the exact Wilson coefficients for light four-quark condensates at our disposal their impact for in-medium situations can be quantified: Within the previously employed Borel window relevant for pseudo-scalar $\bar{q}Q$ and $\bar{Q}q$ excitations the individual contributions to the even part are one to two orders of magnitude smaller than most of the other known terms at saturation density. A similar statement holds for the contributions of the odd part. By comparing to the $\rho$ meson OPE terms we are able to locate the origin of these order of magnitude differences in the $D$ meson contributions:
($i$) the absence of light four-quark terms from hard-gluon diagrams and ($ii$) Wilson coefficients altering more strongly with changing Borel mass due to the non-negligible heavy-quark mass. Despite the small numerical impact of these higher mass dimension contributions, they are required for a profound estimate of the reliable Borel window:
Extending the OPE up to light-quark condensates of mass dimension 6 delivers the bonus to allow for a better determination of the Borel window, because the lower limit is constrained by the highest order OPE terms which must not contribute more than $\sim\!\unit{10}{\%}$ to the OPE \cite{leinweber97}.

Our presentation makes obvious the avenue for improvements: Insertion of the next-to-leading order interaction term into the correlator provides loop corrections to the Wilson coefficients for condensates of lower mass dimension as well as further four-quark condensate contributions with associated diagrams on tree-level. We emphasize the rapidly growing complexity of higher order contributions, especially for in-medium situations. Our evaluation of the Wilson coefficients of leading-order $\alpha_\text{s}$ terms related to light four-quark condensates demonstrates this already. Including, furthermore, condensation of heavy quarks can be accommodated in the present formalism, albeit resulting in cumbersome expressions. Probably new methods are needed for executing the OPE and subsequent evaluation of the sum rules as a complement to lattice QCD methods.

Although the numerical impact of light four-quark condensate terms on the OPE proved to be small, they are structurally important in hadron physics due to their close connection to chiral symmetry. Apart from the chiral condensate which serves as an order parameter of spontaneous chiral symmetry breaking we identify a further chirally odd condensate, $ \displaystyle \langle : \! \bar{q} t^A q \sum_f{ \bar{q}_f \slashed{v} t^A q_f} \! : \rangle $, among the four-quark condensate contributions of the pseudo-scalar $D$ meson sum rule. Chirally odd condensates quantify the difference of chiral partner spectra and can also be studied by Weinberg-type sum rules, proven, e.\,g.\ in \cite{hohlerrapp12,holthohlerrapp13,hohlerrapp14}, as extremely useful when addressing issues of chiral restoration in a strongly interacting medium.

The physics motivation of the present investigation is clearly driven by the contemporary interest in open charm (also bottom) mesons as probes of the hot, strongly interacting medium created in ultra-relativistic heavy-ion collisions (cf.\ \cite{CBMBook,yasuisudoh14,yasuisudoh13b,yasuisudoh13a,blaschke,he,tolos13a,tolos09,kumarmishra11,kumarmishra10} and further references therein). Moreover, the planned experiments of the CBM and PANDA collaborations at FAIR will study charm degrees of freedom in proton and anti-proton induced reactions of nuclei and in heavy-ion collisions as well. For these experimental investigations a solid theoretical basis is mandatory. Among possible approaches with emphasis on medium modifications are QCD sum rules as a method with intimate contact to QCD.

\section*{Acknowledgments}
Enlightening discussions with S.~J.~Brodsky, S.~H.~Lee, S.~Leupold, K.~Morita, U.~Mosel, S.~Narison, R.~Rapp and W.~Weise are gratefully acknowledged. The work is supported by BMBF grant 05P12CRGHE and by the Austrian Science Fund (FWF) under project number P25121-N27.

\appendix*

\section{Calculation of Wilson coefficients of light-quark condensates of mass dimension 6}
\label{sec:calc4Q}

\label{subsec:calcsglu}

For the calculation of Wilson coefficients of light four-quark condensates corresponding to tree-level diagrams containing a soft gluon line we utilize Eq.~\eqref{eq:2qterms}. Using standard OPE techniques -- projection of Dirac indices onto elements of the Clifford algebra as well as the covariant expansion of quark field operators exploiting the Fock-Schwinger gauge -- the light two-quark term $\Pi^{(2)}$ after Fourier transformation reads \cite{hilger11}
\begin{align}
	&\Pi^{(2)} (p) = \sum_a { \frac{1}{4} \sum_{n=0}^{\infty}{ \frac{(-i)^n}{n!} \partial_p^{\vec{\alpha}_n} } } \langle : \! \bar{q} \overset{{}_\leftarrow}{D}_{\vec{\alpha}_n} \Gamma_a \text{Tr}_\text{D}\left[ \Gamma^a \gamma_5 S_Q(p) \gamma_5 \right] q \! : \rangle \, ,
	\label{eq:FT2qterms}
\end{align}
where heavy-quark condensates are neglected. The symbol $ \overset{{}_\leftarrow}{D}_{\vec{\alpha}_n} = \overset{{}_\leftarrow}{D}_{\alpha_1} \ldots \overset{{}_\leftarrow}{D}_{\alpha_n} $ stands for the covariant derivative and an analogous notation for the partial derivative is employed. Quark fields and derivatives thereof are evaluated at $x=0$. The projection of Dirac indices gives the sum over the basis elements $\Gamma_a$ of the Clifford algebra, where $\Gamma_a \in \{ \mathbbm{1}, \gamma_\mu, \sigma_{\mu<\nu}, i\gamma_5\gamma_\mu, \gamma_5 \}$ normalized by the scalar product $\text{Tr}_\text{D} [\Gamma_a \Gamma^b] = 4\delta_a^b$; and we define $\sigma_{\mu\nu} = i (\gamma_\mu \gamma_\nu - g_{\mu\nu})$. Treated in a classical, weak gluonic background field the interaction of the quark is modeled by soft gluon exchange with the QCD ground state and the full propagators $S_Q(p) = \int d^4x\, e^{ipx} S_Q(x,0)$ can be written as
\begin{align}
	S_Q(p) = \sum_{n=0}^\infty S_Q^{(n)}(p)
	\label{eq:pertqprop}
\end{align}
with 
\begin{align}
	S_Q^{(n)}(p) = - S_Q^{(0)}(p) \gamma^\mu \tilde{A}_\mu S_Q^{(n-1)}(p) = - S_Q^{(n-1)}(p) \gamma^\mu \tilde{A}_\mu S_Q^{(0)}(p) \, ,
\end{align}
where $S_Q^{(0)}(p)$ is the free heavy-quark propagator. $\tilde{A}_\mu$ denotes the derivative operator which arises due to the Fourier transform of the perturbative series for the quark propagator in coordinate space from the gluonic background field $A_\mu$:
\begin{align}
	\tilde{A}_\mu = \sum_{n=0}^\infty \tilde{A}^{(n)}_\mu
	\label{eq:backgrA}
\end{align}
with
\begin{subequations}
\begin{align}
	\tilde{A}^{(0)}_\mu &	= i \frac{g}{2} G_{\mu\nu}(0) \partial_p^\nu
	\label{eq:backgrAnull} \\
	\tilde{A}^{(n)}_\mu &	= -g\frac{(-i)^{n+1}}{n!(n+2)} \left[ D_{\vec{\alpha}_n} , G_{\mu\nu}(0) \right]_{(n)} \partial_p^\nu \partial_p^{\vec{\alpha}_n}
	\label{eq:backgrAn} \\
	&	= - g\frac{(-i)^{n+1}}{n!(n+2)} \left[ D_{\alpha_1} , \left[ D_{\alpha_2} , \ldots \left[ D_{\alpha_n} , G_{\mu\nu}(0) \right] \ldots \right] \right] \partial_p^\nu \partial_p^{\alpha_1} \partial_p^{\alpha_2} \ldots \partial_p^{\alpha_n} \, ,
	\label{eq:backgrAnlong}
\end{align}
\end{subequations}
where $G_{\mu\nu}=\frac{i}{g}[D_\mu,D_\nu]=G^A_{\mu\nu} t^A$ is the gluon field strength tensor and $g$ is the coupling. The matrices $t^A$ are the generators of the color group and $A=1,\ldots,N_\text{c}^2-1$ \cite{hilgerbu12}. Throughout this paper, quark propagators constructed from a finite number of terms in Eqs.~\eqref{eq:pertqprop} and \eqref{eq:backgrA} are referred to as perturbative quark propagators.

Light-quark condensate terms in mass dimension 6 enter $\Pi^{(2)}$ with $n=3$ and $S_Q = S^{(0)}_Q$, $n=0$ and $S_Q = S^{(1)}_{Q\,(\tilde{A}^{(1)})}$ as well as $n=1$ and $S_Q = S^{(1)}_{Q\,(\tilde{A}^{(0)})}$ \cite{nov84}, where an additional subscript is introduced to specify the order of the background field expansion. The light-quark condensate contribution reads
\begin{align}
	\Pi_\text{dim6} (p) = \Pi_\text{dim6}^\text{[1]} (p) + \Pi_\text{dim6}^\text{[2]} (p) + \Pi_\text{dim6}^\text{[3]} (p)
\end{align}
with the three terms
\begin{align}
	\Pi_\text{dim6}^\text{[1]} (p) & = \sum_a \frac{1}{4} \frac{(-i)^3}{3!}  \langle : \! \bar{q} \overset{{}_\leftarrow}{D}_\nu \overset{{}_\leftarrow}{D}_\lambda \overset{{}_\leftarrow}{D}_\rho \Gamma_a q \! : \rangle \text{Tr}_\text{D}\left[ \Gamma^a \gamma_5 \partial_p^\nu \partial_p^\lambda \partial_p^\rho S_Q^{(0)} (p) \gamma_5 \right] \, , \\
	\Pi_\text{dim6}^{[2]} (p) & = \sum_a{ \frac{1}{4} } \langle : \! \bar{q} \Gamma_a \text{Tr}_\text{D}\left[ \Gamma^a \gamma_5 S_{Q \, (\tilde{A}^{(1)})}^{(1)} (p) \gamma_5 \right] q \! : \rangle \, , \\
	\Pi_\text{dim6}^{[3]} (p) & = \sum_a{ \frac{1}{4} } \frac{(-i)^1}{1!} \partial_p^\nu \langle : \! \bar{q} \overset{{}_\leftarrow}{D}_\nu \Gamma_a \text{Tr}_\text{D}\left[ \Gamma^a \gamma_5 S_{Q \, (\tilde{A}^{(0)})}^{(1)} (p) \gamma_5 \right] q \! : \rangle \, ,
\end{align}
where $\Pi_\text{dim6}^\text{[1]}$ contains contributions associated  with diagram (d) and $\Pi_\text{dim6}^\text{[2,3]}$ incorporate terms associated with diagram (e) in Fig.~\ref{fig:diagramPiAlpha0(2)}. Using the perturbative quark propagators $ S_{Q\,(\tilde{A}^{(1)})}^{(1)} (p) = - S^{(0)}_Q (p) \gamma^\rho \tilde{A}_\rho^{(1)} S^{(0)}_Q (p) $  and $ S_{Q\,(\tilde{A}^{(0)})}^{(1)} (p) = - S^{(0)}_Q (p) \gamma^\rho \tilde{A}_\rho^{(0)} S^{(0)}_Q (p) $ with $ \tilde{A}_\rho^{(1)} = \frac{g}{3} [ D_\nu , G_{\rho\lambda} ] \partial_p^\lambda \partial_p^\nu $ and $ \tilde{A}_\rho^{(0)} = i \frac{g}{2} G_{\rho\lambda} \partial_p^\lambda $, respectively, one obtains
\begin{align}
	\Pi_\text{dim6}^{[2]} (p) & = - \frac{g}{12} \sum_a \langle : \! \bar{q} \Gamma_a \left[ D_\nu , G_{\rho\lambda} \right] q \! : \rangle \text{Tr}_\text{D}\left[ \Gamma^a \gamma_5 S_Q^{(0)} (p) \gamma^\rho \left( \partial_{p}^\lambda \partial_{p}^\nu S_Q^{(0)} (p) \right) \gamma_5 \right] \, , \\
	\Pi_\text{dim6}^{[3]} (p) & = - \frac{g}{8} \sum_a \langle : \! \bar{q} \overset{{}_\leftarrow}{D}_\nu \Gamma_a G_{\rho\lambda} q \! : \rangle \partial_p^\nu \text{Tr}_\text{D}\left[ \Gamma^a \gamma_5 S_Q^{(0)} (p) \gamma^\rho \left( \partial_{p}^\lambda S_Q^{(0)} (p) \right) \gamma_5 \right] \, .
	\label{eq:missedByHaya}
\end{align}
Subsequently, $\Pi_\text{dim6}^\text{[1,2,3]}$ are treated analogously. Utilizing the identity
\begin{align}
	\partial_p^\mu S_Q^{(0)} (p) = - S_Q^{(0)} (p) \gamma^\mu S_Q^{(0)} (p)
	\label{eq:AblTrick}
\end{align}
and insertion of the free quark propagator $S_Q^{(0)}(p) = (\slashed{p} + m_Q)/(p^2-m_Q^2)$ yields traces of products of Dirac matrices, which stem from elements of the Clifford algebra $\Gamma_a$ as well as free quark propagators and vertex functions. The results of the Dirac trace evaluations are to be contracted with the tensor decompositions of the Gibbs averaged operators $\langle : \! \bar{q} \overset{{}_\leftarrow}{D}_\nu \overset{{}_\leftarrow}{D}_\lambda \overset{{}_\leftarrow}{D}_\rho \Gamma_a q \! : \rangle$, $\langle : \! \bar{q} \Gamma_a \left[ D_\nu , G_{\rho\lambda} \right] q \! : \rangle$ and $\langle : \! \bar{q} \overset{{}_\leftarrow}{D}_\nu \Gamma_a G_{\rho\lambda} q \! : \rangle$.

In order to obtain a continuous transition of the OPE for $T,n \rightarrow 0 $, the in-medium decomposition into Lorentz structures composed of $g_{\mu\nu}$ (metric tensor), $\varepsilon_{\mu\nu\lambda\sigma}$ (Levi-Civita symbol) and $v_\mu$ (medium four-velocity) \cite{buchheim14} requires the separation of vacuum and medium-specific condensate contributions, where the former are present in vacuum while the latter vanish at zero temperature or nucleon density \cite{buchheim14QCD14}. If the (anti-)symmetries within the Lorentz indices of the operators are imposed on the decomposition structures one is able to identify unambiguous medium-specific operators. 

The resulting Gibbs averaged operators can be reduced to condensates of lower mass dimension using the quark equation of motion, or they contain covariant derivatives which can not be eliminated by application of the equation of motion, i.\,e.\ exhibiting light-quark condensation in mass dimension 6. Especially, the combinations $D^\lambda D_\nu D_\lambda$, $[ D^\lambda , G_{\nu\lambda} ]$ and $G_{\nu\lambda} D^\lambda$ incorporate the desired four-quark condensate terms. The first and third terms contain the second combination $[ D^\lambda , G_{\nu\lambda} ]$, which allows for the application of the gluon equation of motion. One obtains
\begin{align}
	\langle : \! \bar{q} \Gamma [ D^\lambda , G_{\nu\lambda} ] q \! :\rangle = g \langle : \! \bar{q}_i \Gamma t^A q_i \sum_f{ \bar{q}_f \gamma_\nu t^A q_f } \! : \rangle \, ,
	\label{eq:conds}
\end{align}
where $ \Gamma \in \{ v^\nu, \gamma^\nu, \slashed{v} v^\nu \} $, because other elements of the Clifford algebra lead to expectation values which are not invariant under time reversal and parity transformations \cite{jin}. The result of these calculations is Eq.~\eqref{eq:result1P} with operators listed in  Tab.~\ref{tab:condoperators} providing the complete light-quark condensate contribution to the OPE of pseudo-scalar $qQ$ mesons in mass \mbox{dimension 6}.

\bibliographystyle{model1a-num-names}
\bibliography{literature}

\end{document}